\documentclass[12pt,letterpaper,leqno]{article} 
\usepackage{amsmath,amsfonts,amssymb,amsthm,enumerate,enumitem,multirow,array,graphicx,lscape,subcaption,color,epstopdf,setspace,mathrsfs,listings,apptools,natbib,algorithmic,algorithm,cite}

\usepackage[letterpaper,margin=1in]{geometry}
\usepackage[bookmarksdepth=2,colorlinks=false]{hyperref}
\usepackage[OT1]{fontenc}
\usepackage{titlesec,titling}
\pretitle{\begin{center}\Large\bfseries}
\posttitle{\par\end{center}}
\preauthor{\begin{center}\lineskip 0.5em\scshape\begin{tabular}[t]{c}}
\postauthor{\end{tabular}\par\end{center}}
\predate{\begin{center}\normalfont}
\postdate{\end{center}\vskip -1em}

\providecommand{\keywords}[1]{\par{\scshape Keywords:} #1}
\providecommand{\jelcodes}[1]{\par{\scshape JEL Codes:} #1}
\titleformat{\section}[hang]{\normalfont\bfseries}{\relax\thesection.~}{0pt}{}[]
\AtAppendix{\titleformat{\section}[block]{\normalfont\scshape\filcenter}{Appendix~\thesection.~}{0pt}{}[]}
\titleformat{\subsection}[runin]{\normalfont\itshape}{\thesubsection.~}{0pt}{}[.]
\AtAppendix{\titleformat{\subsection}[hang]{\normalfont\bfseries}{\relax\thesubsection.~}{0pt}{}[]}
\titleformat{\subsubsection}[runin]{\normalfont\itshape}{\thesubsubsection.~}{0pt}{}[.]
\titlespacing{\section}{\parindent}{1em}{1em}
\titlespacing{\subsection}{\parindent}{1em}{1em}
\titlespacing{\subsubsection}{\parindent}{1em}{1em}
\newtheoremstyle{imsthm}{1em}{1em}{\itshape}{\parindent}{\scshape}{.}{.5em}{}
\theoremstyle{imsthm}
\newtheorem{thm}{Theorem}
\newtheorem{lmm}[thm]{Lemma}
\newtheorem{pro}[thm]{Proposition}

\newtheoremstyle{imsdfn}{1em}{1em}{\normalfont}{\parindent}{\scshape}{.}{.5em}{}
\theoremstyle{imsdfn}

\newtheorem*{rmk}{Remarks}

\AtAppendix{\numberwithin{equation}{section}}
\AtAppendix{\numberwithin{thm}{section}}
\usepackage[font=small,labelfont=sc]{caption}
\usepackage{natbib}
\setlist{noitemsep}

\DeclareMathOperator*{\argmax}{argmax}

\usepackage{tikz}
\usetikzlibrary{calc,fit,positioning}
\usetikzlibrary{decorations.pathreplacing}
\usepackage{epigraph}

\begin{document}

\title{Equilibrium Refinement in Finite Action Evidence Games}
\author{Shaofei Jiang\thanks{The University of Bonn (email: sjiang@uni-bonn.de). I thank V. Bhaskar, William Fuchs, Sven Rady, Maxwell Stinchcombe, Caroline Thomas, Thomas Wiseman, and seminar participants at UT Austin for helpful comments. I gratefully acknowledge funding from the Deutsche Forschungsgemeinschaft (DFG, German Research Foundation) under Germany's Excellence Strategy - GZ 2047/1, Projekt-ID 390685813. Errors are my own.}}
\date{September 18, 2022}

\maketitle

\thispagestyle{empty}

\onehalfspacing

\begin{abstract}

Evidence games study situations where a sender persuades a receiver by selectively disclosing hard evidence about an unknown state of the world. Evidence games often have multiple equilibria. \citet{hart_kremer_perry_2017} propose to focus on truth-leaning equilibria, i.e., perfect Bayesian equilibria where the sender discloses truthfully when indifferent, and the receiver takes off-path disclosure at face value. They show that a truth-leaning equilibrium is an equilibrium of a perturbed game where the sender has an infinitesimal reward for truth-telling. We show that, when the receiver's action space is finite, truth-leaning equilibrium may fail to exist, and it is not equivalent to equilibrium of the perturbed game. To restore existence, we introduce a disturbed game with a small uncertainty about the receiver's payoff. A purifiable truthful equilibrium is the limit of a sequence of truth-leaning equilibria in the disturbed games as the disturbances converge to zero. It exists and features a simple characterization. A truth-leaning equilibrium that is also purifiable truthful is an equilibrium of the perturbed game. Moreover, purifiable truthful equilibria are receiver optimal and give the receiver the same payoff as the optimal deterministic mechanism.

\vskip 1em

\jelcodes{C72, D82, D83}
\end{abstract}

\clearpage
\pagenumbering{arabic}

\section{Introduction}

In many real-life situations, communication relies on hard evidence. For example, a jury's verdict is based on hard evidence presented in the court, rather than exchanges of empty claims. Evidence games study such situations. There is a sender (e.g., a prosecutor), and a receiver (e.g., a jury). The sender has private hard evidence about an unknown state of the world (e.g., whether a defendant is guilty) that she can selectively present to the receiver, and the receiver takes an action (e.g., conviction or acquittal) that is payoff relevant to both players. Full revelation of evidence is often impossible in the presence of conflict of interest between the sender and the receiver--the receiver wants to learn the payoff relevant state and act accordingly, whereas the sender merely wants to induce her preferred receiver action (e.g., convicting the defendant). Therefore, the sender has an incentive to persuade the receiver that a certain state is more likely by partially revealing evidence.

Formally, verifiability of hard evidence is modeled by assuming that the sender's feasible set of disclosure depends on her type (i.e., her evidence), and in this paper, we assume that the sender's type space is ordered.\footnote{See \citet{bull_watson_2004,bull_watson_2007} for discussions on this assumption. Alternatively, \citet{grossman_hart_1980} and \citet{grossman_1981} assume that there is a finite type space, and each sender type can disclose any subset of the type space containing her true type.} That is, some sender types have \emph{more evidence} than others, and the sender can disclose less evidence than she has, hence the feasible set of disclosure is the lower contour set of her type under the ``more evidence'' order. Moreover, the sender's payoff depends only on the receiver's action and not her type or the state of the world. For example, the prosecutor's objective is to convict the defendant. This is not affected by what evidence she has. \emph{In equilibrium}, her chance of convicting the defendant may depend on the evidence she has, because when she has more evidence, there are more ways to present evidence in the court, and thereby she can better persuade the jury.

Evidence games often have multiple (Nash) equilibria. For instance, there is a trivial equilibrium where, regardless of her evidence, the prosecutor presents no evidence to the court, and the jury acquits the defendant regardless of what is presented (this must be optimal on the equilibrium path for the jury if the presumption of innocence is practiced). This is undoubtedly not a sensible prediction of what happens in courtrooms. However, this equilibrium is both perfect \citep{selten_1975} and sequential \citep{kreps_wilson_1982} under mild assumptions.\footnote{Without specifying a complete model, let us assume that the prosecutor is one of four possible types: having no evidence (type \(\emptyset\)), having only evidence supporting conviction (type \(c\)), having only evidence supporting acquittal (type \(a\)), and having both kinds of evidence (type \(ac\)). The type \(ac\) has more evidence than either type \(c\) or type \(a\), who in turn has more evidence than type \(\emptyset\). The jury's payoff is such that they prefer conviction if the prosecutor's type is \(c\), and they strictly prefer acquittal if the prosecutor's type is \(ac\), \(a\), or \(\emptyset\).} Consider a perturbation to the prosecutor's strategy that assigns higher probability on disclosing evidence that supports conviction than on disclosing evidence that supports acquittal, and a perturbation to the jury's strategy such that the probability of convicting the defendant after seeing any evidence is no larger than that after seeing no evidence. As both perturbations diminish, this gives a convergent sequence of \(\varepsilon\)-constrained equilibria in completely mixed strategies. Therefore, the trivial equilibrium is perfect. Similarly, given proper perturbations in the prosecutor's strategy, it is consistent for the jury to hold the belief that, after seeing any evidence, the actual evidence possessed by the prosecutor favors acquitting the defendant. Therefore, this trivial equilibrium is also a sequential equilibrium.

There are also extensive discussions on the value of commitment power in evidence games. That is, whether the receiver can achieve a higher payoff by committing ex ante to a mapping from the sender's disclosure to a distribution over his actions. \citet{glazer_rubinstein_2006} show that there is no value of commitment when the receiver's action is a binary one; \citet{sher_2011} shows the same result when the receiver's payoff is concave in his action, and the receiver's actions can be either finite or continuous.

\citet{hart_kremer_perry_2017} (henceforth HKP) generalize the condition of concavity in \citet{sher_2011} for the case of continuous receiver actions. They focus on the receiver's commitment to a deterministic mechanism. That is, the receiver commits to an action for each possible disclosure, and he cannot randomize over his actions.\footnote{In an earlier version of their paper, \citet{hart_kremer_perry_2015} allow the receiver to randomize and show a stricter condition under which commitment to a stochastic mechanism has no value.} They show that committing to a deterministic mechanism has no value if the receiver chooses an action on the real line, the receiver's expected payoff is a single-peaked function of his action given any distribution of the state, and the sender strictly prefers higher receiver action. Moreover, HKP propose the following equilibrium refinement in evidence games. A \emph{truth-leaning equilibrium} is a perfect Bayesian equilibrium such that\footnote{HKP define truth-leaning equilibrium as a refinement to Nash equilibrium. However, we note that any truth-leaning equilibrium is a perfect Bayesian equilibrium (as is defined in section \ref{sec3}) and sequential equilibrium. We view all solution concepts in the current paper as refinements of perfect Bayesian equilibrium.}
\begin{description}
	\item[(Truth-leaning)] Given the receiver's strategy, the sender discloses her evidence truthfully if doing so is optimal;
	\item[(Off-path beliefs)] The receiver takes any off-path disclosure at face value (i.e., he believes that the sender discloses truthfully).
\end{description}
As is argued in HKP, these conditions follow the straightforward intuition that there is a ``slight inherent advantage'' for the sender to tell the whole truth, and ``there must be good reasons for not telling it.'' Under the assumption that the receiver takes a continuous action, HKP show that a truth-leaning equilibrium exists and is receiver optimal. That is, it gives the receiver the same ex ante payoffs as the optimal deterministic mechanism.

However, in many applications of evidence games, the receiver takes a discrete action. For example, juries choose between conviction and acquittal, banks decide whether or not to grant a loan, and rating agencies rate financial assets into finitely many grades. A part of this paper is to answer the following question: is truth-leaning equilibrium a ``good'' solution concept when the receiver's action set is finite?

The short answer is ``no,'' and one reason is that a truth-leaning equilibrium may fail to exist. Loosely speaking, nonexistence arises because the sender lacks a strict incentive to persuade the receiver.\footnote{In HKP, if a piece of evidence \(e'\) is inherently better than the sender's evidence \(e\) (i.e., the receiver's optimal action knowing that the sender's evidence is \(e'\) is strictly higher than his optimal action knowing that the sender's evidence is \(e\)) and the sender can feasibly disclose \(e'\), then the sender's payoff from any randomization between disclosing \(e'\) and \(e\) is strictly higher than her payoff from disclosing only \(e\), given any Bayesian consistent system of beliefs of the receiver and any sequentially rational receiver strategy. This is not the case when the receiver's action is finite.} Truth-leaning equilibrium also ignores that players often face small payoff uncertainties in evidence games. It is an idealization to assume, for example, that a prosecutor knows perfectly a jury's criteria when making their verdict. Solution concepts that ignore this may lead to unrealistic predictions.

To address these problems, we propose the following solution concept by introducing a small uncertainty (i.e., disturbance) to the receiver's payoff \`{a} la \citet{harsanyi_1973}.\footnote{While Harsanyi's purification theorem has been widely accepted as a leading justification for mixed strategy equilibria, it has also been applied as a refinement in dynamic games (e.g., \citet{bhaskar_mailath_morris_2013,bhaskar_thomas_2019}) and cheap talk games \citep{diehl_kuzmics_2021}. Evidence games are a class of games with nongeneric payoffs, since the sender's action is not payoff relevant. Hence, some equilibria of evidence games are not purifiable.} Suppose that the receiver receives a random private payoff shock associated with each of his actions. In this disturbed game, the sender has a strict incentive to persuade the receiver, and a truth-leaning equilibrium exists.\footnote{For example, the prosecutor does not know how lenient the jury is (i.e., how convinced the jury has to be in order to reach a conviction). However, she knows that after seeing more evidence in favor of conviction, the likelihood that the jury will convict the defendant is strictly higher. Therefore, the prosecutor strictly prefers presenting all evidence that supports conviction.} We define a \emph{purifiable truthful equilibrium} as the limit of a sequence of truth-leaning equilibria in the disturbed games as the disturbances converge to zero. That is, a purifiable truthful equilibrium is a truth-leaning equilibrium of an infinitesimally disturbed game. A purifiable truthful equilibrium always exists and is a perfect Bayesian equilibrium. 

Purifiable truthful equilibria are also receiver optimal. That is, purifiable truthful equilibria maximize the receiver's ex ante payoff among all perfect Bayesian equilibria. And the receiver's purifiable truthful equilibrium payoff is the same as his payoff in the optimal deterministic mechanism. The receiver, however, may achieve a higher payoff than his purifiable truthful equilibrium payoff by committing to a stochastic mechanism. This is akin to the results in HKP.

Another problem of truth-leaning equilibrium in finite evidence games is that it may not follow the intuition that the sender is slightly more advantageous if she discloses truthfully. To formalize this intuition, we revisit the perturbed game in HKP, where the sender receives a small reward if she discloses truthfully, and the sender must disclose truthfully with at least some small probability. We define a \emph{weakly truth-leaning equilibrium} as the limit of a sequence of perfect Bayesian equilibria of the perturbed games as the perturbations converge to zero.\footnote{It is important to make the distinction between a \emph{disturbed game} and a \emph{perturbed game} clear. A disturbed game is a game with a small receiver payoff uncertainty. A perturbed game, as is studied in HKP, is one where both players' payoff functions are public information. Throughout the paper, we refer to them by their respective names.} HKP show that truth-leaning equilibrium is equivalent to weakly truth-leaning equilibrium. When the receiver's action space is finite, however, this equivalence is no longer true. It turns out that purifiability is the missing connection: a weakly truth-leaning equilibrium that is also purifiable truthful is a truth-leaning equilibrium; a truth-leaning equilibrium that is also purifiable truthful is a weakly truth-leaning equilibrium in ``almost all'' (in a precise sense, see Proposition \ref{pro5}) evidence games.

The paper proceeds as follows. Section \ref{sec2} presents a simple example where truth-leaning refinement leads to nonexistence of equilibrium and discusses some other limitations of the existing refinements. We construct the purifiable truthful equilibrium of this example. Section \ref{sec3} models evidence games. Section \ref{sec4} studies purifiable truthful equilibrium and compares various equilibrium refinements of evidence games. The last section concludes. Proofs are in the Appendix.

\section{A Simple Example} \label{sec2}

Every new aircraft design has to be certified by the Federal Aviation Administration (FAA) before any aircraft built according to this design can enter service. Like other innovations, altering the design of an aircraft  often entails high level of risks. The FAA often has to rely on information and test results provided by airplane manufacturers, yet airplane manufactures' disclosure is far from complete.\footnote{For example, design flaws of the battery system on board Boeing's 787 Dreamliners caused two incidents in 2013, which led to the grounding of all aircraft at the time and a redesign of the battery system (see \url{https://www.reuters.com/article/us-boeing-787-battery-idUSKCN0JF35G20141202}). More recently, MCAS, a new flight control software in Boeing's 737 MAX aircraft, caused two deadly crashes within two years of the airliner's first commercial operation. Boeing allegedly did not submit certification documents to FAA detailing changes to the flight control system (see \url{https://www.reuters.com/article/us-boeing-737max-exclusive-idUSKBN2413R6}).}

Imagine an airplane manufacturer (the sender) seeking to get a new aircraft design certified by the FAA (the receiver). The design can be good or bad with equal likelihood. If the design is bad, the aircraft manufacturer has some bad evidence (e.g., mechanical failures during test flights) with probability \(\frac{2}{3}\). Otherwise, the aircraft manufacturer has no evidence. The FAA does not know the quality of the design and chooses to \emph{Approve} (\(a=1\)) or \emph{Reject} (\(a=0\)) the aircraft design based on evidence disclosed by the sender. The disclosure of bad evidence is voluntary and verifiable. That is, disclosing no evidence is always possible, but the airplane manufacturer can disclose bad evidence only if it has bad evidence. The airplane manufacturer's payoff depends only on the FAA's action: it receives 1 if the design is approved and 0 if the design is rejected. The FAA, on the other hand, gains from approving a good design and loses from approving a bad design. Its payoff is 0 if it rejects the design, 1 if it approves a good design, and -2 if it approves a bad design. Hence, the FAA has a cutoff decision rule. If, after observing the disclosed evidence, its posterior belief that the design is good exceeds \(\frac{2}{3}\), its optimal action is \emph{Approve}; if its posterior belief is less than \(\frac{2}{3}\), its optimal action is \emph{Reject}; if its posterior belief is exactly \(\frac{2}{3}\), either action as well as any randomization between the two actions is optimal.

A strategy of the sender describes how it discloses bad evidence. Let \(p\) be the probability that the sender discloses no evidence if it has bad evidence. Since bad evidence fully reveals that the design is bad, the receiver always chooses \emph{Reject} (thus the sender gets 0) after seeing bad evidence. Let \(q\) be the probability that the receiver chooses \emph{Approve} after seeing no evidence. Let \(\mu\) be the receiver's posterior belief that the design is good after seeing no evidence. Since no evidence is disclosed with positive probability, Bayes' rule requires that \(\mu=\frac{3}{4+2p}\).

\subsection{Truth-leaning equilibrium}

It is easy to verify that the game has a continuum of perfect Bayesian equilibria--any \(p\geq\frac{1}{4}\), \(q=0\), and \(\mu=\frac{3}{4+2p}\leq\frac{2}{3}\) constitute an equilibrium. That is, the sender with bad evidence discloses no evidence with at least probability \(\frac{1}{4}\), and the receiver always rejects the new design.

However, there is no truth-leaning equilibrium. Given the receiver's strategy, the sender with bad evidence is indifferent between disclosing no evidence and disclosing truthfully since both actions yield zero payoff. Truth-leaning therefore requires the sender to disclose bad evidence truthfully (i.e., \(p=0\)), which is not satisfied by any perfect Bayesian equilibrium.

\subsection{Purifiable truthful equilibrium}

Suppose that the receiver receives a payoff shock \(\zeta\) for choosing \emph{Approve}, where \(\zeta\) is normally distributed according to \(\mathcal{N}(0,\varepsilon^2)\) and is private information of the receiver (hence the receiver's type). That is, the receiver's payoff from approving a good design is \(1+\zeta\), and that from approving a bad design is \(\zeta-2\). Given any posterior belief \(\mu\), almost all receiver types have a unique optimal action after seeing no evidence, which is \emph{Approve} if \(\mu>\frac{2-\zeta}{3}\) (equivalently, \(\zeta>2-3\mu\)) and \emph{Reject} if \(\mu<\frac{2-\zeta}{3}\) (equivalently, \(\zeta<2-3\mu\)). Hence, in any perfect Bayesian equilibrium of the disturbed game, the design is approved with probability \(\Phi\left(\frac{3\mu-2}{\varepsilon}\right)\) if the sender discloses no evidence, where \(\Phi\) is the cdf of the standard normal distribution. Since this probability is strictly positive for all \(\mu\), the sender strictly prefers disclosing no evidence to disclosing truthfully.

To summarize, let \(q(\zeta)\) denote the probability that the type \(\zeta\) receiver approves the design after observing no evidence. The perfect Bayesian equilibrium of the disturbed game is unique (except for the strategy of a single receiver type), where \(p=1\), \(\mu=\frac{1}{2}\), and \(q(\zeta)=0\) if \(\zeta<\frac{1}{2}\), \(q(\zeta)=1\) if \(\zeta>\frac{1}{2}\). Since the sender strictly prefers disclosing no evidence, this equilibrium is also truth-leaning. In this equilibrium, \emph{Approve} is chosen with probability \(\Phi(-\frac{1}{2\varepsilon})\) after the receiver observes no evidence. That is, the disturbed game has a unique truth-leaning equilibrium outcome: the sender discloses no evidence, and after seeing no evidence, the receiver chooses \emph{Approve} with probability \(\Phi(-\frac{1}{2\varepsilon})\) and believes that the design is good with \(\frac{1}{2}\) probability.

As the disturbance diminishes (i.e., as \(\varepsilon\downarrow 0\)), the unique equilibrium outcome of the disturbed game converges to a perfect Bayesian equilibrium of the original evidence game where the sender discloses no evidence, the receiver always chooses \emph{Reject}, and the receiver's posterior belief on the good design is \(\frac{1}{2}\) after seeing no evidence (i.e., \(p=1\), \(q=0\), \(\mu=\frac{1}{2}\)).

\subsection{Weakly truth-leaning equilibrium}

Consider the following perturbed game. Let \(\varepsilon_1\) and \(\varepsilon_2\) be small positive reals that are common knowledge to the sender and the receiver. The sender receives a reward \(\varepsilon_1\) if it discloses (bad evidence) truthfully, and the sender must disclose truthfully with at least probability \(\varepsilon_2\).

If its posterior belief \(\mu>\frac{2}{3}\), then the receiver has a unique optimal action \emph{Approve} after observing no evidence. Then, for \(\varepsilon_1<1\), the sender strictly prefers disclosing no evidence, so the Bayesian consistent belief is \(\mu=\frac{1}{2}<\frac{2}{3}\). If \(\mu<\frac{2}{3}\), the receiver's unique optimal action is \emph{Reject} after observing no evidence. With the reward for truth-telling, the sender strictly prefers disclosing truthfully, so the Bayesian consistent belief is \(\mu=\frac{3}{4}>\frac{2}{3}\). Hence, the receiver's posterior belief \(\mu=\frac{2}{3}\) in any perfect Bayesian equilibrium of the perturbed game. Indeed, for \(\varepsilon_1<1\) and \(\varepsilon_2\leq\frac{3}{4}\), the perturbed game has a unique perfect Bayesian equilibrium, where \(p=\frac{1}{4}\), \(q=\varepsilon_1\), \(\mu=\frac{2}{3}\).

As \(\varepsilon_1,\varepsilon_2\downarrow0\), the perfect Bayesian equilibrium of the perturbed game converges to a perfect Bayesian equilibrium of the original game, where \(p=\frac{1}{4}\), \(q=0\), \(\mu=\frac{2}{3}\).

\subsection{Discussion}

\begin{figure}[h]
	\centering
	\begin{tikzpicture}[scale=.8]
		\draw (0,0) node[below] {0} -- (3,0) node[below] {1/4};
		\draw (3,0) -- (12,0) node[below] {1};
		\draw [ultra thick, blue] (3,0) -- (12,0);
		\draw[fill = red] (3,0) circle [radius = .1];
		\node[above,align=center] at (3,.4) {Weakly\\truth-\\leaning};
		\draw [->] (3,.5) -- (3,.1);
		\draw[fill = red] (12,0) circle [radius = .1];
		\node[above,align=center] at (12,.4) {Purifiable\\truthful};
		\draw [->] (12,.5) -- (12,.1);
		\draw[decorate, decoration={brace,raise=0}] (3,.2) -- node [above] {PBE} (12,.2);
		\draw[fill = white] (0,0) circle [radius = .1];
		\node[above,align=center] at (0,.4) {Truth-\\leaning\\(non-equilibrium)};
		\draw [->] (0,.5) -- (0,.15);
	\end{tikzpicture}
	\caption{The probability that the sender discloses no evidence when having bad evidence (\(p\))}\label{fig1}
\end{figure}
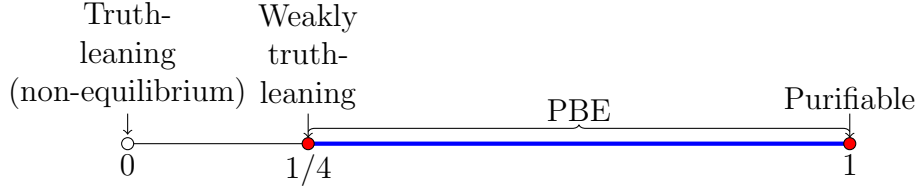

Figure \ref{fig1} illustrates the equilibria of the game. There is a continuum of perfect Bayesian equilibria which differ in the sender's strategy. Among them, the weakly truth-leaning equilibrium maximizes the probability that the sender discloses truthfully. The purifiable truthful equilibrium maximizes the receiver's posterior belief on the good design.

The fact that this simple game does not possess a truth-leaning equilibrium suggests that truth-leaning equilibrium is not an appropriate solution concept for finite evidence games. A more fundamental problem of truth-leaning equilibrium is the discrepancy between the refinement and the intuition behind it. The requirement that the sender weakly prefers disclosing truthfully seemingly stems from the sender having an infinitesimal reward for truth-telling, but in the example, the weakly truth-leaning equilibrium constructed by adding an infinitesimal reward for truth-telling is not the same as imposing the truth-leaning refinement on perfect Bayesian equilibria.\footnote{Recall that HKP show the equivalence of truth-leaning equilibrium and weakly truth-leaning equilibrium in evidence games where the receiver continuously chooses an action, and its payoff function is single-peaked given any belief. In the current example, suppose that the receiver chooses an action \(a\in\mathbb{R}\), and the receiver has quadratic loss utility, i.e., his payoff is \(-(a-x)^2\), where \(x\) is a random variable that equals 0 if the design is bad and 1 if the design is good. The unique truth-leaning equilibrium is as follows. The sender always discloses no evidence, the receiver's belief and action are \(\frac{1}{2}\) after seeing no evidence and 0 after seeing bad evidence. This is also the unique weakly truth-leaning equilibrium.} The following proposition summarizes these observations. The negative result motivates the study of purifiable truthful equilibrium.

\begin{pro} \label{pro6}
	In a finite evidence game, a truth-leaning equilibrium may fail to exist, and a weakly truth-leaning equilibrium may not be truth-leaning.
\end{pro}

Weakly truth-leaning equilibrium exists in finite evidence games, but it also has several shortcomings. A distinctive feature of the weakly truth-leaning equilibrium in the above example is that the receiver is indifferent between choosing \emph{Approve} and \emph{Reject} after seeing no evidence, but it is prescribed to choose only \emph{Reject}. This feature is prevalent and not specific to this example, and it leads to several problems. First, the equilibrium may fail to be perfect. In the example, given any mixed strategy of the receiver, disclosing no evidence is a strictly better response for the sender than disclosing bad evidence truthfully. Hence, the weakly truth-leaning equilibrium where the sender plays a mixed strategy is not a perfect equilibrium.\footnote{We assume that the receiver acts only if no evidence is disclosed. An alternative way to model the example is to let the receiver take an action after each possible disclosure. That is, he has two information sets (one after seeing no evidence, and one after seeing bad evidence) and four pure strategies. In this model, the weakly truth-leaning equilibrium is not a proper equilibrium \citep{myerson_1978} of the normal form game. It is normal form perfect and extensive form perfect and proper.}

Second, weakly truth-leaning equilibrium may not be robust to incomplete receiver payoff information. As is shown above, the sender strictly prefers disclosing no evidence once we introduce a small uncertainty to the receiver's payoff. In defense of weakly truth-leaning equilibrium, the perfect Bayesian equilibrium in every perturbed game where the sender receives a small reward for truth-telling (i.e., \(p=\frac{1}{4}\), \(q=\varepsilon_1\), \(\mu=\frac{2}{3}\)) is robust to incomplete receiver payoff information in our example,\footnote{To see this, consider a disturbed game where: (i) the sender receives \(\varepsilon_1\) if it discloses truthfully; (ii) the sender must disclose truthfully with at least probability \(\varepsilon_2\); (iii) the receiver receives a payoff shock \(\zeta\) distributed according to \(\mathcal{N}(0,\varepsilon^2)\) for choosing \emph{Approve}, which is its private information. For \(\varepsilon_1<\frac{1}{2}\), \(\varepsilon_2<\frac{3}{4}\), and \(\varepsilon<\frac{3-4\varepsilon_2}{6-2\varepsilon_2}\cdot\frac{1}{-\Phi^{-1}(\varepsilon_1)}\), the disturbed game has an essentially unique perfect Bayesian equilibrium (except for the strategy of a single receiver type), where \(p=\frac{9}{4+2\varepsilon\Phi^{-1}(\varepsilon_1)}-2\), \(\mu=\frac{2+\varepsilon\Phi^{-1}(\varepsilon_1)}{3}\), \(q(\zeta)=0\) if \(\zeta<-\varepsilon\Phi^{-1}(\varepsilon_1)\), and \(q(\zeta)=1\) if \(\zeta>-\varepsilon\Phi^{-1}(\varepsilon_1)\). In this equilibrium, the design is approved with probability \(\varepsilon_1\) after the receiver observes no evidence. As \(\varepsilon\downarrow0\), the equilibrium outcome converges to \(p=\frac{1}{4}\), \(q=\varepsilon_1\), \(\mu=\frac{2}{3}\).} but this is not a generic result. In general, a weakly truth-leaning equilibrium may fail to be the limit point of a sequence of equilibria of perturbed games that are robust to incomplete receiver payoff information.

Third, different sequences of perturbations may select different weakly truth-leaning equilibria, and not all sequences of perturbed games have a convergent sequence of perfect Bayesian equilibria as the perturbation goes to zero.

Consider a slight variant to our example, where the sender's bad evidence is either type 1 or type 2 (think about software failures and hardware failures). If the design is bad, the sender has type 1 bad evidence, type 2 bad evidence, and no evidence each with \(\frac{1}{3}\) probability; if the design is good, the sender has no evidence. The sender with a certain type of bad evidence can disclose truthfully or no evidence but cannot disclose the other type of bad evidence. Let \(p_i\) denote the probability that the sender with type \(i\) bad evidence discloses no evidence, \(q\) the probability that the receiver chooses \emph{Approve} after seeing no evidence, and \(\mu\) the receiver's belief that the design is good after seeing no evidence. The game has a continuum of perfect Bayesian equilibria, where \(p_1+p_2\geq\frac{1}{2}\), \(q=0\), and \(\mu=\frac{3}{4+p_1+p_2}\).

Now, let us consider the following perturbed game. Given small positive reals \(\varepsilon_1,\varepsilon_2<1\) and \(\varepsilon_{1|1},\varepsilon_{2|2}\leq\frac{1}{2}\), the sender receives a reward \(\varepsilon_i\) if it truthfully discloses type \(i\) bad evidence, and the sender with type \(i\) bad evidence must disclose truthfully with at least probability \(\varepsilon_{i|i}\). If \(\varepsilon_i<\varepsilon_j\), the unique perfect Bayesian equilibrium is \(p_i=\frac{1}{2}\), \(p_j=0\), \(q=\varepsilon_i\), \(\mu=\frac{2}{3}\). That is, the receiver randomizes between \emph{Approve} and \emph{Reject} after seeing no evidence in order to match the lower reward \(\varepsilon_i\); the sender with type \(i\) bad evidence is indifferent and randomizes between disclosing no evidence and disclosing truthfully, while the sender with type \(j\) bad evidence strictly prefers disclosing truthfully because of the higher reward \(\varepsilon_j\). If \(\varepsilon_1=\varepsilon_2\), there is a continuum of perfect Bayesian equilibria, where \(p_1+p_2=\frac{1}{2}\), \(q=\varepsilon_1=\varepsilon_2\), \(\mu=\frac{2}{3}\). Hence, as \((\varepsilon_1,\varepsilon_{1|1},\varepsilon_2,\varepsilon_{2|2})\to 0\), whether there exists a convergent sequence of perfect Bayesian equilibria depends on the rates of convergence of \(\varepsilon_1\) and \(\varepsilon_2\). If \(\varepsilon_1=\varepsilon_2\) almost always, then any perfect Bayesian equilibrium of the unperturbed game such that \(p_1+p_2=\frac{1}{2}\) is the limit point of a sequence of perfect Bayesian equilibria of the perturbed games. If \(\varepsilon_i\leq\varepsilon_j\) almost always and \(\varepsilon_i<\varepsilon_j\) infinitely often, then the unique weakly truth-leaning equilibrium is \(p_i=\frac{1}{2}\), \(p_j=0\), \(q=0\), and \(\mu=\frac{2}{3}\). If neither case happens, there is no convergent sequence of perfect Bayesian equilibria of the perturbed game. In conclusion, the unperturbed game has a continuum of weakly truth-leaning equilibria, where \(p_1+p_2=\frac{1}{2}\), \(q=0\), \(\mu=\frac{2}{3}\), and different weakly truth-leaning equilibria may be selected by different sets of infinitesimal perturbations.

Purifiable truthful equilibrium is spared from similar problems. For almost all evidence games, purifiable truthful equilibria do not involve the receiver's ``borderline'' beliefs, and any purifiable truthful equilibrium is infinitesimally close to a truth-leaning equilibrium of \emph{any} infinitesimally disturbed game. That is, purifiability does not depend on the selection of disturbances. The normality of the receiver's payoff shock in our example is dispensable. Moreover, the set of purifiable truthful equilibria has a simple structure, and we give a characterization of the set of purifiable truthful equilibria of any evidence game.

\section{The Evidence Game} \label{sec3}

There are two stages. Two players, a sender (she) and a receiver (he), move sequentially. At the outset of the game, a state of the world \(\omega\in\{G,B\}\) is realized with probability \(\pi_0\in(0,1)\) on \(\omega=G\). Neither player observes the realized state \(\omega\),\footnote{Since the sender's payoff is independent of the realized state, it does not change our analysis if the realized state is known to the sender.} and the prior \(\pi_0\) is common knowledge. In the first stage, the sender observes a piece of hard evidence \(e\in E\) and discloses \(m\in E\) to the receiver, where \(E\) is a finite set of evidence. In the second stage, the receiver observes the disclosed evidence \(m\) and chooses an action \(a\in A\), where \(A=\{a_1<a_2<\dots<a_K\}\) is a finite subset of the real line with \(K\geq 2\).

\subsection{Evidence and disclosure}

Let \(F_G\) and \(F_B\) be two distributions over the set of evidence \(E\). The sender's evidence \(e\) is a random draw from either \(F_G\) or \(F_B\), depending on the realized state. If \(\omega=G\), \(e\) is drawn from distribution \(F_G\); if \(\omega=B\), it is drawn from distribution \(F_B\).

Disclosure is verifiable. That is, the set of evidence that the sender can feasibly disclose depends on the evidence she has (in contrast, in a signaling game, the sender chooses from the same set of signals regardless of her type). Throughout the paper, we maintain the following assumptions that are standard in the literature:
\begin{description}
	\item[(Reflexivity)] The sender can always truthfully disclose her evidence \(e\);
	\item[(Transitivity)] If the sender can disclose \(e'\) when she has evidence \(e\), and she can disclose \(e''\) when she has evidence \(e'\), then she can disclose \(e''\) if she has evidence \(e\).
\end{description}
Under these assumptions, we can represent the ``disclosure rule'' as a preorder \(\precsim\) on \(E\). Disclosing \(m\) is feasible given evidence \(e\) if and only if \(m\precsim e\), and the feasible set of disclosure given a piece of evidence \(e\) is its lower contour set \(\{m\in E:m\precsim e\}\), denoted \(LC(e)\).

\subsection{Payoffs}

The receiver's payoff \(u_R(a,\omega)\) depends on both his action and the realized state of the world (but not the true evidence or the disclosed evidence), and the receiver maximizes his expected payoff.\footnote{Equivalently, one can assume that \(E\subset\mathbb{R}\) and that the receiver's payoff \(u_R(a,e)\) is linear in the sender's evidence \(e\).} We assume that the receiver's payoff function satisfies the following assumption:\footnote{If no receiver action is dominated (i.e., every action is the receiver's unique optimal action at some belief), the assumption of increasing differences is equivalent to the assumption of single-peakedness in HKP, i.e., for all \(\mu\in[0,1]\), there exists a single-peaked function \(f_\mu:\mathbb{R}\to\mathbb{R}\) such that \(\mu u_R(a,G)+(1-\mu)u_R(a,B)=f_\mu(a)\) for all \(a\in A\).}
\begin{description}
	\item[(Increasing differences)] \(u_R(a,G)-u_R(a,B)\) is strictly increasing in \(a\).
\end{description}
Under this assumption, the receiver wants to match the state of the world. That is, his optimal action is weakly increasing in his posterior belief that the state is good. More precisely, given \(\mu\in[0,1]\), the solution to the receiver's maximization problem
\begin{equation*}
	\phi(\mu) = \argmax_{a\in A} \mu u_R(a,G) + (1-\mu)u_R(a,B)
\end{equation*}
is upper hemicontinuous and weakly increasing in \(\mu\).\footnote{Throughout the paper, we say a correspondence \(\phi:[0,1]\rightrightarrows A\) is \emph{weakly increasing} if \(a_i\leq a_j\) for all \(\mu_i<\mu_j\), \(a_i\in\phi(\mu_i)\), and \(a_j\in\phi(\mu_j)\).}

The sender's payoff equals the receiver's action, i.e., \(u_S(a,\omega)=a\). Given the assumption on the receiver's payoff, the sender has a weak incentive to persuade the receiver that the state is good. Notice that the evidence \(e\), the disclosed evidence \(m\), and the realized state \(\omega\) are payoff irrelevant to the sender.

An evidence game is a tuple \(\mathcal{G}=\langle\pi_0,(E,\precsim),F_G,F_B,A,u_R\rangle\).

\subsection{Strategies and perfect Bayesian equilibrium}

A strategy of the sender is \(\sigma:E\to\Delta(E)\) such that \(supp(\sigma(\cdot|e))\subset LC(e)\), a strategy of the receiver is \(\rho:E\to\Delta(A)\), and a system of beliefs of the receiver is \(\mu:E\to[0,1]\), where \(\mu(m)\) denotes the receiver's posterior belief that the state is good after observing \(m\).

A \emph{perfect Bayesian equilibrium} of \(\mathcal{G}\) is a collection of the sender's strategy, the receiver's strategy, and the receiver's system of belief \((\sigma,\rho,\mu)\) such that:
\begin{description}
	\item[(Sender optimality)] Given \(\rho\),
	\begin{equation*}
		supp(\sigma(\cdot|e)) \subset \argmax_{m\precsim e}\sum_{a\in A}a\cdot\rho(a|m)
	\end{equation*}
	for all \(e\in E\);
	\item[(Receiver optimality)] Given \(\mu\),
	\begin{equation*}
		supp(\rho(\cdot|m)) \subset \phi(\mu(m))
	\end{equation*}
	for all \(m\in E\);
	\item[(Bayesian consistency)] For all on-path disclosure \(m\in\bigcup_{e\in E}supp(\sigma(\cdot|e))\),
	\begin{equation*}
		\mu(m)=\frac{\sum_{e\in E}\sigma(m|e)F_G(e)\pi_0}{\sum_{e\in E}\sigma(m|e)[F_G(e)\pi_0+F_B(e)(1-\pi_0)]}.
	\end{equation*}
\end{description}

\section{Refinements of Perfect Bayesian Equilibrium} \label{sec4}

Sections \ref{sec4.1} through \ref{sec4.3} study truth-leaning equilibrium, purifiable truthful equilibrium, and weakly truth-leaning equilibrium. Section \ref{sec4.4} shows the relationship between these refinements. Section \ref{sec5} shows that purifiable truthful equilibria are receiver optimal, and that there is no value of committing to a deterministic mechanism.

\subsection{Truth-leaning equilibrium} \label{sec4.1}

A \emph{truth-leaning equilibrium} of \(\mathcal{G}\) is a perfect Bayesian equilibrium \((\sigma,\rho,\mu)\) such that:
\begin{description}
	\item[(Truth-leaning)] Given \(\rho\),
	\begin{equation*}
		e \in \argmax_{m\precsim e}\sum_{a\in A}a\cdot\rho(a|m) \Rightarrow \sigma(e|e) = 1;
	\end{equation*}
	\item[(Off-path beliefs)] For all off-path disclosure \(m\), \(\mu(m)=\nu(m)\), where
	\begin{equation*}
		\nu(m) = \frac{F_G(m)\pi_0}{F_G(m)\pi_0+F_B(m)(1-\pi_0)}.
	\end{equation*}
\end{description}

As the example in Section \ref{sec2} shows, a truth-leaning equilibrium may not exist. The following proposition shows that nonexistence happens extensively. Fix an evidence structure and vary only the receiver's payoffs. Unless ``more evidence'' implies ``better evidence'' (i.e., \(\nu\) is weakly increasing), there is a positive measure of evidence games in which a truth-leaning equilibrium does not exist.

\begin{pro} \label{pro8}
	Fix \(\pi_0,(E,\precsim),F_G,F_B\), and \(A\). Let \(\mathscr{G}\) be the set of all evidence games with prior \(\pi_0\), evidence space \((E,\precsim)\), distributions of evidence \(F_G\) and \(F_B\), and receiver action space \(A\). Identify \(\mathscr{G}\) with a subset of \(\mathbb{R}^{2K}\) by the bijection
    \begin{equation*}
    	\langle\pi_0,(E,\precsim),F_G,F_B,A,u_R\rangle\mapsto\{u_R(a,G),u_R(a,B)\}_{a\in A}.
    \end{equation*}
    If \(\nu:(E,\precsim)\to[0,1]\) is weakly increasing, then every evidence game in \(\mathscr{G}\) has a truth-leaning equilibrium, and in all truth-leaning equilibria, the sender discloses truthfully. If \(\nu\) is not weakly increasing, a truth-leaning equilibrium does not exist in a positive measure of evidence games in \(\mathscr{G}\).
\end{pro}

\subsection{Disturbed games and purifiable truthful equilibrium} \label{sec4.2}

A disturbed game is where the receiver has a private payoff shock (i.e., his type) \(\zeta:A\to\mathbb{R}\). The receiver has type dependent payoff \(v_R(a,\omega|\zeta)=u_R(a,\omega)+\zeta(a)\). We identify the set of the receiver's types with \(\mathbb{R}^K\), where \(K=|A|\) is the number of available receiver actions. Let \(\eta\) be a distribution over \(\mathbb{R}^K\) that has full support and is absolutely continuous with respect to the Lebesgue measure.\footnote{The assumption that the disturbance has full support is dispensable. For every purifiable truthful equilibrium, there exists a sequence of disturbances which assign positive probability to finitely many payoff shocks, and a sequence of truth-leaning equilibria of the disturbed games that converges to the purifiable truthful equilibrium.} Denote by \(\mathcal{G}_R(\eta)\) the disturbed game where the receiver's type is distributed according to \(\eta\).

In the disturbed game, a strategy of the sender is \(\sigma:E\to\Delta(E)\) such that \(supp(\sigma(\cdot|e))\subset LC(e)\), a strategy of the receiver in \(\mathcal{G}_R(\eta)\) is \(r:E\times\mathbb{R}^K\to\Delta(A)\), and a system of beliefs of the receiver is \(\mu:E\to[0,1]\), where \(\mu(m)\) is the receiver's posterior belief that the state is good after observing \(m\).\footnote{For the purpose of finding truth-leaning equilibria, it is without loss to assume that the receiver's belief is independent of his type, since on-path beliefs are determined by Bayes' rule, and off-path beliefs are determined by the refinement.} Given any strategy of the receiver \(r\), let \(\rho:E\to\Delta(A)\) be the induced distributions over the receiver's actions. That is,
\begin{equation*}
	\rho(a|m) = \int r(a|m,\zeta)\eta(d\zeta)
\end{equation*}
is the probability that the receiver takes action \(a\) after \(m\) is disclosed. We shall also use the shorthand notation and write this as \(\rho=\langle r,\eta\rangle\).

A \emph{truth-leaning equilibrium} of \(\mathcal{G}_R(\eta)\) is a tuple \((\sigma,r,\mu)\) such that:
\begin{description}
	\item[(Receiver optimality in disturbed games)] Given \(\mu\),
	\begin{equation*}
		supp(r(\cdot|m,\zeta)) \subset \tau(\mu(m),\zeta)
	\end{equation*}
	for all \(m\in E\) and \(\zeta\in\mathbb{R}^K\), where \(\tau(\bar{\mu},\zeta)\subset A\) is the solution to the type \(\zeta\) receiver's problem given posterior belief \(\bar{\mu}\in[0,1]\) on the good state, i.e.,
	\begin{equation*}
		\tau(\bar{\mu},\zeta) = \argmax_{a\in A} \bar{\mu} u_R(a,G)+(1-\bar{\mu})u_R(a,B)+\zeta(a);
	\end{equation*}
	\item \textbf{(Sender optimality)}, \textbf{(Bayesian consistency)}, \textbf{(Truth-leaning)}, and \textbf{(Off-path beliefs)}, as are defined above for the original game \(\mathcal{G}\).
\end{description}
If \((\sigma,r,\mu)\) is a truth-leaning equilibrium, we say \((\sigma,\rho,\mu)\) is a \emph{truth-leaning equilibrium outcome} of \(\mathcal{G}_R(\eta)\).

In any disturbed game, the sender has a strict incentive to persuade the receiver. That is, from the sender's perspective, the expected value of the receiver's optimal action is strictly increasing in his posterior belief. Therefore, a truth-leaning equilibrium exists in any disturbed game, and it is equivalent to a weakly truth-leaning equilibrium. Moreover, a truth-leaning equilibrium is essentially unique, and the receiver's equilibrium system of beliefs depends only on the evidence space \((E,\precsim)\) and the distributions \(F_G\) and \(F_B\). Specifically, it is independent of the disturbance \(\eta\).\footnote{Indeed, as is shown in Appendix \ref{appendix.a1}, it is independent of the receiver's payoff function \(u_R\) in the undisturbed evidence game, but for the purpose of Lemma \ref{lmm1}, we consider only disturbed games of a fixed evidence game.} Hence, the receiver's equilibrium system of beliefs is the same across \emph{all} truth-leaning equilibria of \emph{all} disturbed games. In fact, the set of truth-leaning equilibria is the same in all disturbed games.

\begin{lmm} \label{lmm1}
	A truth-leaning equilibrium exists in all disturbed games. Moreover, there exist a closed and convex set \(\Sigma^\star\subset\Delta(E)^E\) and a system of beliefs of the receiver \(\mu^\star\) such that for all disturbed games \(\mathcal{G}_R(\eta)\), \((\sigma,r,\mu)\) is a truth-leaning equilibrium of \(\mathcal{G}_R(\eta)\) if and only if \(\sigma\in\Sigma^\star\), \(\mu=\mu^\star\), and \(supp(r(\cdot|m,\zeta))\subset\tau(\mu(m),\zeta)\) for all \(m\in E\) and \(\zeta\in\mathbb{R}^K\).
\end{lmm}

We define a purifiable truthful equilibrium as the limit point of a sequence of truth-leaning equilibrium outcomes of the disturbed games as the payoff uncertainty goes to zero. Formally, a \emph{purifiable truthful equilibrium} of \(\mathcal{G}\) is a tuple \((\sigma,\rho,\mu)\) such that there exists a sequence of disturbances \(\{\eta^n\}_{n=1}^\infty\), and for each \(\eta^n\), a truth-leaning equilibrium outcome \((\sigma^n,\rho^n,\mu^n)\) of \(\mathcal{G}_R(\eta^n)\) such that \(\eta^n\) converges weakly to the point mass at \(0\), denoted \(\eta^n\xrightarrow{w}\delta_0\), and \((\sigma^n,\rho^n,\mu^n)\to(\sigma,\rho,\mu)\).

By Lemma \ref{lmm1}, it is easy to see that a purifiable truthful equilibrium exists, and in any purifiable truthful equilibrium, \(\sigma\in\Sigma^\star\) and \(\mu=\mu^\star\). Since the receiver's problem in any disturbed game depends only on his type and his posterior belief, the receiver's action after seeing the disclosed evidence in any purifiable truthful equilibrium should depend only on his posterior belief. That is, if two pieces of evidence \(m\) and \(m'\) are such that \(\mu^\star(m)=\mu^\star(m')\), then \(\rho(\cdot|m)=\rho(\cdot|m')\) in any purifiable truthful equilibrium. Conversely, any perfect Bayesian equilibrium satisfying these conditions is a purifiable truthful equilibrium.

\begin{thm} \label{pro2}
	A purifiable truthful equilibrium exists and is a perfect Bayesian equilibrium. Moreover, \((\sigma,\rho,\mu)\) is a purifiable truthful equilibrium if and only if \(\sigma\in\Sigma^\star\), \(\mu=\mu^\star\), \(supp(\rho(\cdot|m))\subset\phi(\mu(m))\) for all \(m\in E\), and \(\mu(m)=\mu(m')\Rightarrow\rho(\cdot|m)=\rho(\cdot|m')\).
\end{thm}

Theorem \ref{pro2} shows that the set of purifiable truthful equilibria is a connected set in all evidence games. If \(\phi(\mu^\star(m))\) is a singleton for all \(m\in E\), then the receiver's purifiable truthful equilibrium strategy is unique and is a pure strategy (i.e., \(\rho(a|m)=\mathbf{1}_{a\in\phi(\mu^\star(m))}\)), and all purifiable truthful equilibria differ only on the sender's strategies. Since the sender's action is payoff irrelevant, all purifiable truthful equilibria have the same payoff relevant outcome in the sense that the joint distribution of \((a,\omega)\) is the same. Moreover, any purifiable truthful equilibrium can be approached using arbitrary disturbances (see the remarks in Appendix \ref{appendix.a2}). That is, for all purifiable truthful equilibria \((\sigma,\rho,\mu)\) and all sequences of disturbances \(\eta^n\xrightarrow{w}\delta_0\), a sequence of truth-leaning equilibrium outcomes \((\sigma^n,\rho^n,\mu^n)\) of the disturbed games \(\mathcal{G}_R(\eta^n)\) converges to \((\sigma,\rho,\mu)\). In Appendix \ref{appendix.a5}, we show that this is generic: for almost all evidence games, the receiver's purifiable truthful equilibrium strategy is unique and in pure strategies; consequently, any purifiable truthful equilibrium is infinitesimally close to a truth-leaning equilibrium of any infinitesimally disturbed game.

Because the disclosed evidence is payoff irrelevant, these results do not follow the standard results for ``generic'' extensive form games. Since the sender can have a continuum of equilibrium strategies, and different strategies correspond to different joint distributions of \((e,m)\), there is a continuum of purifiable truthful equilibrium outcomes defined as distributions over terminal nodes identified by \((a,e,m,\omega)\) in evidence games. In contrast, for almost all finite extensive form games, the set of Nash equilibrium outcomes is finite \citep{kreps_wilson_1982}. It is also worth mentioning that purifiable truthful equilibria may not be regular (consider the associated normal form game and apply the definition of regularity by \citet{van_damme_1996}) even in generic evidence games. For example, in the purifiable truthful equilibrium of the example in Section \ref{sec2}, given that the receiver always chooses \emph{Reject}, the sender is indifferent between any value of \(p\). The equilibrium \(p=1\), \(q=0\) is therefore irregular.

For nongeneric evidence games, i.e., where \(\phi(\mu^\star(m))\) is not a singleton for some \(m\in E\), there is a continuum of the receiver's purifiable truthful equilibrium strategies. A given purifiable truthful equilibrium may be the limit point of truth-leaning equilibrium outcomes only for some sequences of disturbed games, and not all sequences of disturbed games have a convergent sequence of truth-leaning equilibrium outcomes. For example, consider a slight variant of the example in Section \ref{sec2} where the receiver's payoff from approving a bad design is -1 (instead of -2). As a result, the receiver's belief threshold is \(\frac{1}{2}\). There exists a continuum of purifiable truthful equilibria, where \(p=1\), \(q\in[0,1]\), \(\mu=\frac{1}{2}\). Specifically, there exists a purifiable truthful equilibrium in which the receiver chooses \emph{Approve} (\(a=1\)) and \emph{Reject} (\(a=0\)) with equal probability after seeing no evidence, i.e., \(q=\frac{1}{2}\). But in order to approach this equilibrium using truth-leaning equilibrium outcomes of the disturbed games, the sequence of disturbances \(\{\eta^n\}_{n=1}^\infty\) must satisfy \(\eta^n(\{\zeta(1)>\zeta(0)\})\to\frac{1}{2}\). That is, the probability that the receiver has a strict incentive to choose \emph{Approve} at belief \(\frac{1}{2}\) must converge to \(\frac{1}{2}\), equating the probability that the receiver chooses \emph{Approve} in the intended purifiable truthful equilibrium.

\subsection{Perturbed games and weakly truth-leaning equilibrium} \label{sec4.3}

Let \(\varepsilon=\{\varepsilon_e,\varepsilon_{e|e}\}_{e\in E}\) be a collection of positive real numbers. The perturbed game \(\mathcal{G}_S(\varepsilon)\), as is defined in HKP, is an evidence game where the sender who has evidence \(e\) receives an extra payoff \(\varepsilon_e\) if she discloses truthfully, and she must disclose truthfully with at least probability \(\varepsilon_{e|e}\). That is, the sender's payoff is \(v_S(a,e,m)=a+\varepsilon_e\mathbf{1}_{e=m}\), and a strategy of the sender is \(\sigma:E\to\Delta(E)\) such that \(supp(\rho(\cdot|e))\subset LC(e)\) and \(\sigma(e|e)\geq\varepsilon_{e|e}\) for all \(e\).

A \emph{perfect Bayesian equilibrium} of \(\mathcal{G}_S(\varepsilon)\) is a collection of the sender's strategy, the receiver's strategy, and the receiver's system of beliefs \((\sigma,\rho,\mu)\) such that:
\begin{description}
	\item[(Sender optimality)] Given \(\rho\),
	\begin{equation*}
		\sigma(m|e) > 0 \Rightarrow m\in\argmax_{m\precsim e} \sum_{a\in A} v_S(a,e,m)\cdot\rho(a|m)
	\end{equation*}
	for all \(e\) and \(m\neq e\);
	\item \textbf{(Receiver optimality)} and \textbf{(Bayesian consistency)}, as are defined for \(\mathcal{G}\).
\end{description}

A \emph{weakly truth-leaning equilibrium} of \(\mathcal{G}\) is a tuple \((\sigma,\rho,\mu)\) such that there exists a sequence of perturbations \(\{\varepsilon^n\}_{n=1}^\infty\) and for each \(\varepsilon^n\), a PBE \((\sigma^n,\rho^n,\mu^n)\) of \(\mathcal{G}_S(\varepsilon^n)\) such that \(\varepsilon^n\to0\), and \((\sigma^n,\rho^n,\mu^n)\to(\sigma,\rho,\mu)\).

\begin{pro} \label{pro3}
	A weakly truth-leaning equilibrium exists and is a perfect Bayesian equilibrium.
\end{pro}

Unlike purifiable truthful equilibria, weakly truth-leaning equilibria often involve ``borderline'' receiver beliefs where the receiver is indifferent between two actions (see the remarks in Appendix \ref{appendix.a4}). Consequently, they may not be proper, and different sequences of perturbations may select different weakly truth-leaning equilibria, as is the case in the example in Section \ref{sec2}.

\subsection{Relationship between truth-leaning, weakly truth-leaning, and purifiable truthful equilibria} \label{sec4.4}

HKP show that truth-leaning equilibrium and weakly truth-leaning equilibrium are equivalent in a class of evidence games where the receiver continuously chooses an action on the real line. The leading example in Section \ref{sec2} shows that this is not true for finite evidence games. It turns out that the equivalence can be restored for purifiable truthful equilibria. On the one hand, if a purifiable truthful equilibrium is weakly truth-leaning, it is also a truth-leaning equilibrium. On the other hand, in almost all evidence games, a purifiable truthful equilibrium that is also truth-leaning is a weakly truth-leaning equilibrium.

\begin{pro} \label{pro4}
	If a purifiable truthful equilibrium is weakly truth-leaning, it is also a truth-leaning equilibrium.
\end{pro}

\begin{pro} \label{pro5}
	Fix \(\pi_0,(E,\precsim),F_G,F_B\), and \(A\). Let \(\mathscr{G}\subset\mathbb{R}^{2K}\) be the set of all evidence games with prior \(\pi_0\), evidence space \((E,\precsim)\), distributions of evidence \(F_G\) and \(F_B\), and receiver action space \(A\). Let \(\mathscr{N}\subset\mathscr{G}\) be the set of evidence games that have a purifiable truthful equilibrium that is truth-leaning but not weakly truth-leaning. \(\mathscr{N}\) has Lebesgue measure zero.
\end{pro}

For nongeneric games, a purifiable truthful equilibrium that is truth-leaning equilibrium need not be weakly truth-leaning. Consider again the example presented in Section \ref{sec4.2} where the receiver's belief threshold is \(\frac{1}{2}\). There exists a continuum of truth-leaning equilibria, where \(p=1\), \(q>0\), and \(\mu=\frac{1}{2}\). That is, the sender always discloses no evidence, and the receiver approves the design with positive probability so that the sender's incentive to disclose no evidence is strict. All truth-leaning equilibria are purifiable truthful. However, there is a unique weakly truth-leaning equilibrium, where \(p=1\), \(q=1\), and \(\mu=\frac{1}{2}\). That is, the receiver must choose \emph{Approve} after seeing no evidence. This is because, in every perturbed game, the sender cannot report no evidence with probability one, the receiver's posterior belief is therefore strictly higher than the belief threshold \(\frac{1}{2}\), and he strictly prefers choosing \emph{Approve} after seeing no evidence. This example is not generic, since the receiver is indifferent between \emph{Approve} and \emph{Reject} after seeing no evidence.

The example in Section \ref{sec2} also shows that purifiable truthful equilibrium and weakly truth-leaning equilibrium do not imply each other, and neither implies truth-leaning equilibrium. To complete this part, we now show, using a variant of the example in Section \ref{sec2}, that an equilibrium that is both truth-leaning and weakly truth-leaning can fail to be purifiable truthful. Suppose that we alter the distribution of the sender's evidence when the design is good such that the sender has bad evidence and no evidence with equal probability. The distribution when the design is bad remains unchanged. The game has a unique truth-leaning equilibrium, where \(p=0\), \(q=0\), \(\mu=\frac{3}{5}\).\footnote{Although bad evidence is not fully revealing of the state, it reveals that the sender has bad evidence. Therefore, the receiver's posterior belief on the good design is \(\frac{3}{7}\), and the receiver's unique optimal action is \emph{Reject}. Hence, we can describe an equilibrium of the game using \(p,q,\mu\), as are defined in Section \ref{sec2}.} That is, the sender discloses truthfully, the receiver always rejects the design, and the receiver's belief on the good design is \(\frac{3}{5}\) after seeing no evidence. Notice that this is also the unique weakly truth-leaning equilibrium of the game. However, it is not a purifiable truthful equilibrium. In the unique purifiable truthful equilibrium, the sender always discloses no evidence, the receiver always rejects the project, and its belief on the good design is \(\frac{1}{2}\) after seeing no evidence (i.e., \(p=1,q=0,\mu=\frac{1}{2}\)).

\subsection{Receiver optimality and value of commitment} \label{sec5}

A justification for truth-leaning equilibria in HKP is that they are receiver optimal. Moreover, HKP show that optimal deterministic mechanisms give the receiver the same expected payoff as truth-leaning equilibria.\footnote{A deterministic mechanism is an action plan \(\gamma:E\to A\). In the mechanism design problem, the receiver moves first and publicly chooses a mechanism \(\gamma\). The sender privately observes her evidence \(e\) and optimally choose \(m\precsim e\). The action \(\gamma(m)\) is then taken, and payoffs are realized. An optimal deterministic mechanism maximizes the receiver's expected payoff among all deterministic mechanisms.} Nonexistence of truth-leaning equilibrium prevents us from claiming the same in finite evidence games. The following proposition shows that purifiable truthful equilibria have these desired properties.

\begin{pro} \label{pro7}
	Purifiable truthful equilibria are receiver optimal. The receiver's purifiable truthful equilibrium payoff equals his payoff in the optimal deterministic mechanism.
\end{pro}

To better illustrate this result, consider modifying the example in Section \ref{sec2} as follows. The receiver's belief threshold is \(\frac{1}{3}\), and the sender has conclusive good evidence with probability \(\frac{2}{3}\) if the state is good. That is, the sender can have (conclusive) good evidence, (conclusive) bad evidence, or no evidence. In the unique purifiable truthful equilibrium, the sender truthfully discloses good evidence and never discloses bad evidence, and the receiver chooses \emph{Accept} if and only if good evidence is disclosed. Besides this purifiable truthful equilibrium, there exists a continuum of perfect Bayesian equilibria, where the sender type with bad evidence discloses no evidence, the sender type with good evidence discloses no evidence with at least \(\frac{1}{4}\) probability, and the receiver always chooses \emph{Accept} on the equilibrium path. Clearly, the purifiable truthful equilibrium gives the receiver a strictly higher expected payoff than any other perfect Bayesian equilibrium.

\citet{sher_2011} shows that if the receiver's payoff is a concave function in his action, even stochastic mechanisms cannot give the receiver a higher payoff than his payoff in the receiver optimal equilibrium. Our assumption of increasing differences is weaker, and the receiver may achieve a higher payoff than his purifiable truthful equilibrium payoff by committing to a stochastic mechanism. As an example, consider adding a third receiver action, \emph{Use at own risk} (\(a=\frac{1}{2}\)), to the example in Section \ref{sec2}. If this action is chosen, the sender's payoff is \(\frac{1}{2}\), and the receiver's payoff is \(x\) if the design is good and \(-x\) if the design is bad, where \(0<x<\frac{1}{4}\). Notice that the receiver's payoff function is not concave in his action in the good state. Let \(q(0),q(\frac{1}{2})\), and \(q(1)\) denote the receiver's probability of choosing \emph{Reject}, \emph{Use at own risk}, and \emph{Approve} after seeing no evidence, respectively. And again, \(p\) is the sender's probability of disclosing no evidence when having bad evidence, and \(\mu\) is the receiver's posterior belief that the design is good after seeing no evidence. This game has a continuum of perfect Bayesian equilibria, where \(p=1\), \(q(0)+q(\frac{1}{2})=1\), and \(\mu=\frac{1}{2}\). That is, the sender always discloses no evidence, the receiver chooses \emph{Reject}, \emph{Use at own risk}, or randomizes between these two actions after seeing no evidence, and the receiver's posterior belief on the good design is \(\frac{1}{2}\) after seeing no evidence.\footnote{All perfect Bayesian equilibria are purifiable truthful, and all perfect Bayesian equilibria where \(q(\frac{1}{2})>0\) are truth-leaning. The unique weakly truth-leaning equilibrium has \(q(\frac{1}{2})=1\).} In every equilibrium, the receiver's ex ante expected payoff is zero. The receiver can achieve a positive payoff by committing to a stochastic mechanism. Suppose that it commits to choosing \emph{Use at own risk} after seeing bad evidence and randomizing over \emph{Reject} and \emph{Approve} with equal probability after seeing no evidence. It is then optimal for the sender to disclose bad evidence truthfully, since it receives \(\frac{1}{2}\) regardless. As the sender discloses evidence truthfully, the receiver's ex ante expected payoff is \(\frac{1}{12}-\frac{1}{3}x>0\).

\section{Conclusion}

HKP propose truth-leaning equilibrium as a solution concept in evidence games. The intuition is that the sender may find it slightly more advantageous to disclose evidence truthfully when being indifferent. This paper points out two problems of applying this solution concept to finite evidence games. First, it may fail to exist. Second, it may not agree with the intuition that the sender receives an infinitesimal reward for truth-telling. That is, truth-leaning equilibrium is not equivalent to weakly truth-leaning equilibrium in finite evidence games.

We propose a simple solution to restore existence by adding a small payoff uncertainty to the receiver. In the disturbed game, the sender is as if she faces a single receiver whom she has strict incentive to persuade, and therefore, a truth-leaning equilibrium exists. A  purifiable truthful equilibrium is a truth-leaning equilibrium in an infinitesimally disturbed game. We show that a purifiable truthful equilibrium always exists and characterize the set of purifiable truthful equilibria.

We also show the equivalence between truth-leaning and weakly truth-leaning for purifiable truthful equilibria. If a purifiable truthful equilibrium is weakly truth-leaning, it is also a truth-leaning equilibrium. Conversely, in almost all finite evidence games, a purifiable truthful equilibrium that is also truth-leaning is a weakly truth-leaning equilibrium.

Finally, we show that purifiable truthful equilibria are receiver optimal, and the receiver cannot achieve a higher payoff by committing to a deterministic mechanism.

\appendix

\section{Proofs}

\subsection{Proof of Proposition \ref{pro8}}

\begin{proof}
	Fix an evidence game \(\mathcal{G}\in\mathscr{G}\), and suppose that \(\nu\) is weakly increasing. Since the receiver's optimal action correspondence \(\phi\) is also weakly increasing, it is easy to verify that \((\sigma,\rho,\mu)\) such that \(\sigma(e|e)=1\), \(\mu(m)=\nu(m)\), and \(\rho(a|m)=1\) if and only if \(a=\max\phi(\nu(m))\) for all \(e,m\in E\) is a truth-leaning equilibrium of \(\mathcal{G}\).

	We now show that in all truth-leaning equilibria, the sender discloses truthfully. To obtain a contradiction, suppose that there exists a truth-leaning equilibrium \((\sigma,\rho,\mu)\) such that \(\sigma(m|e)>0\) for some \(m\neq e\). Then the receiver's expected action after seeing \(m\) must exceed that after seeing \(e\). Therefore, \(\mu(m)>\mu(e)\). Since \(m\) is an on-path message, \(\mu(m)\) is by Bayes' rule and is a convex combination of \(\{\nu(e')|m\precsim e'\}\). Since \(\nu\) is increasing, \(\mu(m)\leq\nu(m)\). On the other hand, \(e\) is off-path, because \(m\) is feasible and gives a strictly higher payoff than \(e\) for all sender types who can feasibly disclose \(e\). Hence, \(\mu(e)=\nu(e)\). But \(\nu(m)\leq\nu(e)\), which is a contradiction to \(\mu(m)>\mu(e)\). Therefore, \(\sigma(e|e)=1\) for all \(e\in E\) in every truth-leaning equilibrium of every game in \(\mathscr{G}\).

	Let \(\nu_1<\nu_2<\dots<\nu_N\) be elements of \(\nu(E)\) and \(E_i=\nu^{-1}(\nu_i)\). If \(\nu\) is not weakly increasing, \(N\geq 2\), and we can define
	\begin{equation*}
		\bar{\nu}_i = \max_{m\in\cup_{j<i}E_j}\frac{F_G(E_i\cup\{m\})\pi_0}{F_G(E_i\cup\{m\})\pi_0+F_B(E_i\cup\{m\})(1-\pi_0)} < \nu_i
	\end{equation*}
	for all \(i\geq 2\). Let \(i^\star\) be the largest \(i\) such that there exist \(m\in E_i\) and \(e\in\cup_{j<i}E_j\), and \(m\precsim e\). The existence of \(i^\star\) is guaranteed by the assumption that \(\nu\) is not weakly increasing. Let \(M^\star=\{m\in E_{i^\star}:\exists e\in\cup_{j<i^\star}E_j, m\precsim e\}\), \(E^\star=\{e\in\cup_{j<i^\star}E_j:\exists m\in E_{i^\star}, m\precsim e\}\).

	We now define a set of games where no truth-leaning equilibrium exists. Let \(\mathscr{S}\) be the set of all evidence games in \(\mathscr{G}\) such that \(\phi(\bar{\nu}_{i^\star})=\{a_1\}\) and \(\phi(\nu_{i^\star})=\{a_K\}\). That is, the highest action is uniquely optimal at belief \(\nu_{i^\star}\), and the lowest action is uniquely optimal at belief \(\bar{\nu}_{i^\star}\). The set \(\mathscr{S}\) has positive Lebesgue measure, and no game in \(\mathscr{S}\) has a truth-leaning equilibrium. Suppose that, contrary to our claim, \((\sigma,\rho,\mu)\) is a truth-leaning equilibrium of some \(\mathcal{G}\in\mathscr{S}\). Then the sender's expected payoff from disclosing any \(m\in M^\star\) must exceed \(a_1\). Otherwise, the receiver chooses the lowest action \(a_1\) after seeing some \(m\in M^\star\), so no other sender type would disclose \(m\) with positive probability in a truth-leaning equilibrium. Hence, \(\mu(m)=\nu(m)=\nu_{i^\star}\), and \(a_K\) is the receiver's unique optimal action after seeing \(m\), a contradiction. Therefore, in the truth-leaning equilibrium, all sender types in \(M^\star\cup E^\star\) disclose with probability one messages in \(M^\star\); all sender types in \(\cup_{j\geq i^\star}E_j\setminus M^\star\) disclose truthfully and get \(a_K\); all sender types in \(\cup_{j<i^\star}E_j\setminus E^\star\) disclose truthfully and get \(a_1\). By Bayes' rule, at some \(m\in M^\star\),
	\begin{equation*}
		\mu(m)\leq\frac{F_G(E^\star\cup M^\star)\pi_0}{F_G(E^\star\cup M^\star)\pi_0+F_B(E^\star\cup M^\star)(1-\pi_0)}\leq\bar{\nu}_{i^\star}.
	\end{equation*}
	Hence, after seeing \(m\), \(a_1\) is the receiver's unique optimal action. This is a contradiction to the statement that the sender's expected payoff from disclosing \(m\) exceeds \(a_1\). Therefore, a truth-leaning equilibrium does not exist in any evidence game in \(\mathscr{S}\).
\end{proof}

\subsection{Proof of Lemma \ref{lmm1}} \label{appendix.a1}

\begin{proof}
	Given any posterior belief \(\bar{\mu}\in[0,1]\) on the good state, two actions \(a_i\) and \(a_j\) are both optimal for receiver type \(\zeta\) only if \(\zeta(a_j)-\zeta(a_i)=\bar{\mu}[u_R(a_i,G)-u_R(a_j,G)]+(1-\bar{\mu})[u_R(a_i,B)-u_R(a_j,B)]\). By assumption, this is true only for an \(\eta\)-null set of \(\zeta\). Hence, \(\tau(\bar{\mu},\cdot)\) is \(\eta\)-a.e. a singleton set. This allows us to define
	\begin{equation*}
		\varphi(\bar{\mu}) = \int\sup\tau(\bar{\mu},\zeta)\eta(d\zeta) = \int\inf\tau(\bar{\mu},\zeta)\eta(d\zeta).
	\end{equation*}
	In any equilibrium \((\sigma,r,\mu)\) of the disturbed game, \(\varphi(\mu(m))\) is the sender's expected payoff if she discloses \(m\).

	Moreover, \(\tau(\cdot,\zeta)\) is weakly increasing for all \(\zeta\in\mathbb{R}^K\). Let \(\mu_i<\mu_j\), \(a_i\in\tau(\mu_i,\zeta)\), and \(a_j\in\tau(\mu_j,\zeta)\). Then
	\begin{align*}
		&\mu_i u_R(a_i,G)+(1-\mu_i)u_R(a_i,B)+\zeta(a_i) \geq \mu_i u_R(a_j,G)+(1-\mu_i)u_R(a_j,B)+\zeta(a_j), \\
		&\mu_j u_R(a_j,G)+(1-\mu_j)u_R(a_j,B)+\zeta(a_j) \geq \mu_j u_R(a_i,G)+(1-\mu_j)u_R(a_i,B)+\zeta(a_i).
	\end{align*}
	Hence,
	\begin{equation} \label{eq.a1}
		(\mu_j-\mu_i)[u_R(a_j,G)-u_R(a_j,B)] \geq (\mu_j-\mu_i)[u_R(a_i,G)-u_R(a_i,B)].
	\end{equation}
	Since \(u_R(a,G)-u_R(a,B)\) is strictly increasing in \(a\), (\ref{eq.a1}) implies that \(a_j\geq a_i\).

	Therefore, \(\varphi:[0,1]\to\mathbb{R}\) is strictly increasing. Suppose that, contrary to the claim, there exist \(\mu_i<\mu_j\) such that \(\varphi(\mu_i)=\varphi(\mu_j)\). Then, for all \(\zeta\) except for on a \(\eta\)-null set, \(\tau(\mu_i,\zeta)=\tau(\mu_j,\zeta)\). This is true only if \(u_R(a,G)-u_R(a,B)\) is constant across all \(a\in A\), which contradicts the assumption of increasing differences.

	Now consider an auxiliary evidence game \(\mathcal{G}(\varphi)\) without receiver type, where the receiver chooses an action in \(\mathbb{R}\), and given any posterior belief \(\mu\in[0,1]\), he has a unique optimal action \(\varphi(\mu)\).\footnote{There are different ways to define the receiver's payoff \(\tilde{u}_R:\mathbb{R}\times\{G,B\}\to\mathbb{R}\) in the auxiliary evidence game. Let us assume that \(\tilde{u}_R(a,\omega)=\int v_R(\tau(\varphi^{-1}(a),\zeta),\omega|\zeta)\eta(d\zeta)\) for all \(a\in\varphi([0,1])\).} This is the standard setup in \citet{jiang_2019}. We are to establish a duality between truth-leaning equilibria of \(\mathcal{G}_R(\eta)\) and truth-leaning equilibria of \(\mathcal{G}(\varphi)\).

	Let \((\hat{\sigma},\hat{\mathbf{a}},\hat{\mu})\) be a truth-leaning equilibrium of \(\mathcal{G}(\varphi)\).\footnote{\(\hat{\mathbf{a}}:E\to\mathbb{R}\) is a pure strategy of the receiver. Since given any posterior belief \(\mu\), the receiver has a unique optimal action \(\varphi(\mu)\) in the auxiliary game, he uses a pure strategy such that \(\hat{\mathbf{a}}=\varphi\circ\hat{\mu}\) in any equilibrium of \(\mathcal{G}(\varphi)\).} Let \(r:E\times\mathbb{R}^K\to\Delta(A)\) be such that \(supp(r(\cdot|m,\zeta))\subset\tau(\hat{\mu}(m),\zeta)\) for all \(m\in E\) and \(\zeta\in\mathbb{R}^K\). We are to show that \((\hat{\sigma},r,\hat{\mu})\) is a truth-leaning equilibrium of \(\mathcal{G}_R(\eta)\). By construction, it satisfies receiver optimality, Bayesian consistency, and the condition on off-path beliefs. We only need to verify sender optimality and truth-leaning. Since \(\tau(\hat{\mu}(m),\cdot)\) is \(\eta\)-a.e. a singleton for all \(m\), \(r(a|m,\zeta)=\mathbf{1}_{a\in\tau(\hat{\mu}(m),\zeta)}\) for all \(m,a\), and almost all \(\zeta\). Hence, with slight abuse of notation, \(\sum_{a\in A}a\cdot r(a|m,\zeta)=\tau(\hat{\mu}(m),\zeta)\) for all \(m\) and almost all \(\zeta\). Integrating over \(\zeta\) on both sides, \(\sum_{a\in A}a\cdot\rho(a|m)=\varphi(\hat{\mu}(m))=\hat{\mathbf{a}}(m)\). That is, the sender's problem given \(\hat{\mathbf{a}}\) in \(\mathcal{G}(\varphi)\) is the same as the sender's problem given \(r\) in \(\mathcal{G}_R(\eta)\). Since \((\hat{\sigma},\hat{\rho},\hat{\mu})\) is sender optimal and truth-leaning, \((\hat{\sigma},r,\hat{\mu})\) is therefore also sender optimal and truth-leaning.

	Conversely, let \((\hat{\sigma},\hat{r},\hat{\mu})\) be a truth-leaning equilibrium of \(\mathcal{G}_R(\eta)\), and define \(\mathbf{a}=\varphi\circ\hat{\mu}\). It is easy to see that \((\hat{\sigma},\mathbf{a},\hat{\mu})\) is a truth-leaning equilibrium of \(\mathcal{G}(\varphi)\).

	We now use the characterization of truth-leaning equilibria of the auxiliary evidence game in \citet{jiang_2019}. Fixing a finite evidence space \((E,\precsim)\) and distributions \(F_G\) and \(F_B\), a truth-leaning equilibrium of \(\mathcal{G}(\varphi)\) exists for all strictly increasing \(\varphi:[0,1]\to\mathbb{R}\). Moreover, there exists a system of beliefs \(\mu^\star:E\to[0,1]\) such that for all strictly increasing \(\varphi\), \((\sigma,\mathbf{a},\mu)\) is a truth-leaning equilibrium of \(\mathcal{G}(\varphi)\) if and only if \(\mu=\mu^\star\), \(\mathbf{a}=\varphi\circ\mu\), \(\sigma(e|e)=\mathbf{1}_{\mu(e)\leq\nu(e)}\), and
	\begin{equation}\label{eq.a2}
		\mu(m) = \min\left\{\nu(m),\frac{\sum_{e\in E}\sigma(m|e)F_G(e)\pi_0}{\sum_{e\in E}\sigma(m|e)[F_G(e)\pi_0+F_B(e)(1-\pi_0)]}\right\}
	\end{equation}
	for all \(m\in E\). Notice that (\ref{eq.a2}) defines a continuous mapping \(f:\Delta(E)^E\to[0,1]^E,\sigma\mapsto\mu\). Therefore, \(\Sigma^\star=f^{-1}(\mu^\star)\) is a closed subset of \(\Delta(E)^E\), and \((\sigma,\mathbf{a},\mu)\) is a truth-leaning equilibrium of \(\mathcal{G}(\varphi)\) if and only if \(\sigma\in\Sigma^\star\), \(\mu=\mu^\star\), and \(\mathbf{a}=\varphi\circ\mu\). It is also easy to verify that \(\Sigma^\star\) is convex. Hence, by the above duality, for all disturbances \(\eta\), \((\sigma,r,\mu)\) is a truth-leaning equilibrium of the disturbed game \(\mathcal{G}_R(\eta)\) if and only if \(\sigma\in\Sigma^\star\), \(\mu=\mu^\star\), and \(supp(r(\cdot|m,\zeta))\subset\tau(\mu(m),\zeta)\) for all \(m\in E\) and \(\zeta\in\mathbb{R}^K\).
\end{proof}

\subsection{Proof of Theorem \ref{pro2}} \label{appendix.a2}

\begin{proof}
	The first statement is implied by the second statement, since \(\mu^\star\) is Bayesian consistent with any sender's strategy \(\sigma\in\Sigma^\star\) by Lemma \ref{lmm1}, and \(\phi\) is nonempty-valued.

	For the ``only if '' part of the second statement, let \((\sigma,\rho,\mu)\) be a purifiable truthful equilibrium. There exists a sequence of disturbances \(\eta^n\xrightarrow{w}\delta_0\) and for each \(\eta^n\), a truth-leaning equilibrium \((\sigma^n,r^n,\mu^n)\) of \(\mathcal{G}_R(\eta^n)\) such that \((\sigma^n,\rho^n,\mu^n)\to(\sigma,\rho,\mu)\), where \(\rho^n=\langle r^n,\eta^n\rangle\). By Lemma \ref{lmm1}, \(\sigma^n\in\Sigma^\star\) for all \(n\), and since \(\Sigma^\star\) is closed, \(\sigma\in\Sigma^\star\). Additionally, \(\mu^n=\mu^\star\) for all \(n\), so \(\mu=\mu^\star\). Fix any \(m\in E\) and \(a\in A\) such that \(a\notin\phi(\mu(m))=\tau(\mu(m),0)\). Since \(\tau\) is upper hemicontinuous in \(\zeta\), there exists a neighborhood \(\mathcal{U}\) of \(0\) in \(\mathbb{R}^K\) such that \(a\notin\tau(\mu(m),\zeta)\) for all \(\zeta\in\mathcal{U}\). By receiver optimality, \(r^n(a|m,\zeta)=0\) for all \(n\) and \(\zeta\in\mathcal{U}\). Hence, as \(\eta^n\xrightarrow{w}\delta_0\), \(\rho^n(a|m)=\int r^n(a|m,\zeta)\eta^n(d\zeta)\to 0\). That is, \(a\notin supp(\rho(\cdot|m))\). Lastly, let \(m,m'\in E\) be such that \(\mu(m)=\mu(m')\). Since \(\tau(\mu(m),\zeta)=\tau(\mu(m'),\zeta)\) for all \(\zeta\), \(r^n(a|m,\zeta)=r^n(a|m',\zeta)\) for all \(n\), \(a\), and almost all \(\zeta\). Therefore, \(\rho^n(\cdot|m)=\rho^n(\cdot|m')\) for all \(n\), so their limits also coincide, i.e., \(\rho(\cdot|m)=\rho(\cdot|m')\).

	For the ``if'' part of the second statement, let \((\sigma,\rho,\mu)\) be such that \(\sigma\in\Sigma^\star\), \(\mu=\mu^\star\), \(supp(\rho(\cdot|m))\subset\phi(\mu(m))\) for all \(m\in E\), and \(\mu(m)=\mu(m')\Rightarrow\rho(\cdot|m)=\rho(\cdot|m')\). We are to show that it is a purifiable truthful equilibrium by construction. Let \(\mu_1<\mu_2<\dots<\mu_N\) be elements of \(\mu(E)\), i.e., all posterior beliefs of the receiver after seeing some disclosed evidence. Since \(\tau\) is upper hemicontinuous in \(\zeta\), there exists \(r>0\) such that \(\tau(\mu_i,\zeta)\subset\phi(\mu_i)\) for all \(i\) and all \(\zeta\in B_r(0)\), where \(B_r(0)\) is the open ball of radius \(r\) around \(0\) in \(\mathbb{R}^K\). For each \(\alpha=(\alpha_1,\alpha_2,\dots,\alpha_N)\in\times_{i=1}^N\phi(\mu_i)\), let \(V_\alpha\) be the set of \(\zeta\in B_r(0)\) such that \(\tau(\mu_i,\zeta)=\{\alpha_i\}\) for all \(i=1,2,\dots,N\). That is, \(V_\alpha\) is the set of receiver types who have a unique optimal action \(\alpha_i\) given each belief \(\mu_i\). Notice that \(V_\alpha\)'s are pairwise disjoint, \(\bigcup_\alpha\overline{V_\alpha}=\overline{B_r(0)}\), and \(\lambda\zeta\in V_\alpha\) for all \(\zeta\in V_\alpha\) and \(\lambda\in(0,1)\). Moreover, each \(V_\alpha\) has positive Lebesgue measure.\footnote{Since \(\tau\) is upper hemicontinuous in \(\zeta\), we only need to show that \(V_\alpha\) is nonempty for all \(\alpha\). Notice that \(\tau(\mu_i,\zeta)=\{\alpha_i\}\) if and only if \(\zeta(\alpha_i)>\zeta(a')\) for all \(a'\in\phi(\mu_i)\), \(a'\neq \alpha_i\). The assumption of increasing differences guarantees that \(\phi(\mu_i)\)'s are ranked, and \(\phi(\mu_i)\cap\phi(\mu_j)\) is either empty or a singleton. Therefore, all inequalities can be simultaneously satisfied for all \(i\). That is, there exists \(\zeta\in\mathbb{R}^K\) such that \(\tau(\mu_i,\zeta)=\{\alpha_i\}\) for all \(i\). Hence, \(V_\alpha\) is nonempty.} Let \(q^n\to\rho\) be a sequence such that \(supp( q^n(\cdot|m))=\phi(\mu(m))\) for all \(m\in E\), and \(\mu(m)=\mu(m')\Rightarrow q^n(\cdot|m)=q^n(\cdot|m')\). By abuse of notation, we write \(q^n(a|m)\) as \(q^n(a,\mu(m))\), and let \(x_\alpha^n=\Pi_{i=1}^N q^n(\alpha_i,\mu_i)\). For all \(\alpha\) and all \(n\), \(x_\alpha^n>0\). Therefore, for each \(n\), we can define a distribution \(\eta^n\) over \(\mathbb{R}^K\) with full support and absolutely continuous with respect to the Lebesgue measure such that \(\eta^n\left(\frac{1}{n}V_\alpha\right)=\frac{n-1}{n}x_\alpha^n\) for all \(\alpha\), where \(\frac{1}{n}V_\alpha=\{\zeta:n\zeta\in V_\alpha\}\) is a subset of \(V_\alpha\). That is, \(\eta^n\) assigns increasingly large probability on the open ball of radius \(\frac{1}{n}\) around 0. By construction, \(\eta^n\xrightarrow{w}\delta_0\). Let \(r\) be any receiver strategy in the disturbed games such that \(supp(r(\cdot|m,\zeta))\subset\tau(\mu(m),\zeta)\) for all \(m\in E\) and \(\zeta\in\mathbb{R}^K\). By Lemma \ref{lmm1}, \((\sigma,r,\mu)\) is a truth-leaning equilibrium of \(\mathcal{G}_R(\eta^n)\). Let \((\sigma,\rho^n,\mu)\) be the associated equilibrium outcome. Notice that \(\rho^n(a|m)=\int r(a|m,\zeta)\eta^n(d\zeta)\) is bounded from below by \(\frac{n-1}{n}q^n(a|m)\) and from above by \(\frac{n-1}{n}q^n(a|m)+\frac{1}{n}\), and recall that \(q^n\to\rho\). Hence, \(\rho^n\to\rho\), and \((\sigma,\rho,\mu)\) is a purifiable truthful equilibrium.
\end{proof}

\begin{rmk}
	The above proof implies that, if \(\phi(\mu^\star(m))\) is a singleton for all \(m\in E\), there exists a sequence of truth-leaning equilibrium outcomes of the disturbed games \((\sigma^n,\rho^n,\mu^n)\) that converges to \((\sigma,\rho,\mu)\) for all purifiable truthful equilibria \((\sigma,\rho,\mu)\) and all disturbances \(\eta^n\xrightarrow{w}\delta_0\). Let \(r\) be any receiver strategy in the disturbed games such that \(supp(r(\cdot|m,\zeta))\subset\tau(\mu^\star(m),\zeta)\) for all \(m\in E\) and \(\zeta\in\mathbb{R}^K\). By Lemma \ref{lmm1}, \((\sigma,r,\mu)\) is a truth-leaning equilibrium of all disturbed games \(\mathcal{G}_R(\eta^n)\). Since \(\tau\) is upper hemicontinuous in \(\zeta\), and \(\phi(\mu^\star(m))\) is a singleton for all \(m\in E\), \(r(a|m,\cdot)\) is constant on a small neighborhood of \(0\) in \(\mathbb{R}^K\) for all \(a\in A\) and \(m\in E\). Hence, \(\rho^n(a|m)\to r(a|m,0)=\rho(a|m)\) for all \(a\in A\) and \(m\in E\).
\end{rmk}

\subsection{Proof of Proposition \ref{pro3}}

\begin{proof}
	The proof works similarly as the proof of Proposition 1 in HKP despite differences in our settings. First, observe that a perfect Bayesian equilibrium exists in every perturbed game \(\mathcal{G}_S(\varepsilon)\). The set of sender strategies in the perturbed game \(\Sigma\subset \Delta(E)^E\) and the set of receiver strategies \(\Delta(A)^E\) are convex and compact. Given a strategy of the receiver, the set of the sender's strategies that are optimal is closed and nonempty. This yields an upper hemicontinuous best response correspondence of the sender \(\Gamma_S:\Delta(A)^E\rightrightarrows\Sigma\). Given a sender strategy \(\sigma\), since all evidence is disclosed with positive probability, there is a unique Bayesian consistent system of beliefs \(\mu^\sigma\), and the mapping \(\sigma\mapsto\mu^\sigma\) is continuous. Since the solution to the receiver's optimality problem \(\phi\) is upper hemicontinuous, we have an upper hemicontinuous best response correspondence of the receiver \(\Gamma_R:\Sigma\rightrightarrows\Delta(A)^E\) such that \(\Gamma_R(\sigma)=\times_{m\in E}\Delta(\phi(\mu^\sigma(m)))\). Then by the Kakutani fixed point-theorem, there exists \(\sigma,\rho\) such that \(\sigma\in\Gamma_S(\rho)\) and \(\rho\in\Gamma_R(\sigma)\). That is, the perturbed game has a Nash equilibrium. The Nash equilibrium \((\sigma,\rho)\) paired with the system of beliefs \(\mu^\sigma\) consists of a perfect Bayesian equilibrium of the perturbed game.

	Since the set of the sender's strategies \(\{\sigma:supp(\sigma(\cdot|e)\subset LC(e)\}\subset\Delta(E)^E\), the set of the receiver's strategies \(\Delta(A)^E\), and the set of systems of beliefs \([0,1]^E\) are compact, any sequence of perfect Bayesian equilibria of perturbed games \(\{(\sigma^n,\rho^n,\mu^n)\}_{n=1}^\infty\) has a convergent subsequence. Hence, a weakly truth-leaning equilibrium exists. It is straightforward to verify that any weakly truth-leaning equilibrium is also a perfect Bayesian equilibrium.
\end{proof}

\subsection{Proof of Proposition \ref{pro4}} \label{appendix.a4}

\begin{proof}
	Let \((\sigma,\rho,\mu)\) be a weakly truth-leaning equilibrium that is also purifiable truthful. We show that (i) if \(\sigma(e|e)>0\), then \(\sigma(e|e)=1\), and (ii) if \(\sigma(e|e)=0\), then \(e\notin\argmax_{m\precsim e}\sum_{a\in A}a\cdot\rho(a|m)\), and \(\mu(e)=\nu(e)\). It then follows that \((\sigma,\rho,\mu)\) is also a truth-leaning equilibrium.

	The first claim is due to \((\sigma,\rho,\mu)\) being a purifiable truthful equilibrium. Let \(\eta^n\xrightarrow{w}\delta_0\), and \((\sigma^n,\rho^n,\mu^n)\to(\sigma,\rho,\mu)\) be such that \((\sigma^n,\rho^n,\mu^n)\) is a truth-leaning outcome of \(\mathcal{G}_R(\eta^n)\) for all \(n\). If \(\sigma(e|e)>0\), then there exists \(N\) such that \(\sigma^n(e|e)>0\) for all \(n\geq N\). However, \((\sigma^n,\rho^n,\mu^n)\) is truth-leaning, so \(\sigma^n(e|e)=1\) for all \(n\geq N\). Therefore, \(\sigma(e|e)=1\).

	The second claim is due to weakly truth-leaning. Let \(\varepsilon^n\to 0\), and \((\sigma^n,\rho^n,\mu^n)\to(\sigma,\rho,\mu)\) be such that \((\sigma^n,\rho^n,\mu^n)\) is a perfect Bayesian equilibrium of \(\mathcal{G}_S(\varepsilon^n)\) for all \(n\). If \(\sigma(e|e)=0\), \(e\notin\argmax_{m\precsim e}\sum_{a\in A}a\cdot\rho(a|m)\). Otherwise, for all \(n\), \(e\) is the unique maximizer to the sender's problem in \(\mathcal{G}_S(\varepsilon^n)\), so \(\sigma^n(e|e)=1\), and \(\sigma^n\not\to\sigma\). Hence, for all \(e'\succsim e\) and all \(n\), \(\sigma^n(e|e')=0\). By Bayes' rule, \(\mu^n(e)=\nu(e)\) for all \(n\). Therefore, \(\mu(e)=\nu(e)\).
\end{proof}

\begin{rmk}
	In a similar fashion to the proof above, we can show that if a weakly truth-leaning equilibrium \((\sigma,\rho,\mu)\) does not involve ``borderline'' receiver beliefs, i.e., \(\phi(\mu(m))\) is a singleton for all \(m\in E\), then it is a truth-leaning equilibrium. Hence, in all evidence games where a truth-leaning equilibrium does not exist, all weakly truth-leaning equilibria involve ``borderline'' receiver beliefs where the receiver is indifferent between two actions.
\end{rmk}

\subsection{Proof of Proposition \ref{pro5}} \label{appendix.a5}

\begin{proof}
	Notice that the receiver's system of beliefs \(\mu^\star\) is the same across all purifiable truthful equilibria of all games in \(\mathscr{G}\). Moreover, given any two actions \(a_i,a_j\) and a belief \(\mu\in[0,1]\), the receiver is indifferent between actions \(a_i\) and \(a_j\) at \(\mu\) if and only if \(u(a_i,G),u(a_i,B),u(a_j,G),u(a_j,B)\) are on a hyperplane in \(\mathbb{R}^4\). Therefore, the receiver is indifferent between two actions at some belief \(\mu^\star(m)\) only on a Lebesgue null set of \(\mathcal{G}\). We are to show that, if \(\phi(\mu^\star(m))\) is a singleton for all \(m\in E\), then a truth-leaning equilibria that is also purifiable truthful is weakly truth-leaning. This concludes that \(\mathscr{N}\) has Lebesgue measure zero.

	Let \(\mathcal{G}\in\mathscr{G}\) be such that \(\phi(\mu^\star(m))\) is a singleton for all \(m\), and \((\sigma,\rho,\mu^\star)\) a truth-leaning equilibrium of \(\mathcal{G}\) that is also purifiable truthful. Given any perturbation \(\varepsilon=\{\varepsilon_e,\varepsilon_{e|e}\}_{e\in E}\), we define \((\sigma_\varepsilon,\rho_\varepsilon,\mu_\varepsilon)\) as follows:
	\begin{enumerate}[label=(\arabic*)]
		\item \(\sigma_\varepsilon(e|e)=1\) if \(\sigma(e|e)=1\);
		\item \(\sigma_\varepsilon(e|e)=\varepsilon_{e|e}\), and \(\sigma_\varepsilon(m|e)=(1-\varepsilon_{e|e})\sigma(m|e)\) for all \(m\neq e\) if \(\sigma(e|e)=0\);
		\item \(\rho_\varepsilon=\rho\);
		\item \(\mu_\varepsilon\) is by Bayes' rule, i.e.,
		\begin{equation*}
			\mu_\varepsilon(m)=\frac{\sum_{e\in E}\sigma_\varepsilon(m|e)F_G(e)\pi_0}{\sum_{e\in E}\sigma_\varepsilon(m|e)[F_G(e)\pi_0+F_B(e)(1-\pi_0)]}.
		\end{equation*}
	\end{enumerate}
	
	For sufficiently small \(\varepsilon\), \((\sigma_\varepsilon,\rho_\varepsilon,\mu_\varepsilon)\) is a perfect Bayesian equilibrium of \(\mathcal{G}(\varepsilon)\). Sender optimality is satisfied if
	\begin{equation*}
		\varepsilon_e < \max_{m\in E} \sum_{a\in A} a[\rho(a|m)-\rho(a|e)]
	\end{equation*}
	for all \(e\in E\) such that \(\sigma(e|e)=0\). For all \(m\in E\), since \(\phi\) is upper hemicontinuous and \(\mu^\star(m)\) is a singleton for all \(m\), there exists \(\delta>0\) such that \(\phi(\mu)=\phi(\mu^\star(m))\) for all \(m\) and all \(\mu\in[0,1]\) such that \(|\mu-\mu^\star(m)|<\delta\). Since \(\mu_\varepsilon\to\mu^\star\), when \(\varepsilon\) is sufficiently small, \(\rho_\varepsilon(a|m)=\rho(a|m)=\mathbf{1}_{a=\phi(\mu_\varepsilon(m))}=\mathbf{1}_{a=\phi(\mu^\star(m))}\) for all \(m\in E\). That is, receiver optimality is satisfied. By construction, it is also Bayesian consistent, and \((\sigma_\varepsilon,\rho_\varepsilon,\mu_\varepsilon)\to(\sigma,\rho,\mu)\) for any sequence \(\varepsilon\to0\). Therefore, \((\sigma,\rho,\mu)\) is a weakly truth-leaning equilibrium.
\end{proof}

\subsection{Proof of Proposition \ref{pro7}}

\begin{proof}
	We first show that every receiver optimal perfect Bayesian equilibrium coexists with a payoff equivalent purifiable truthful equilibrium, in the sense that the receiver's ex ante payoff and the each sender type's interim payoff are the same in the two equilibria.

	Let \(\hat{v}(\mu)=\max_{a\in A}\mu u_R(a,G)+(1-\mu)u_R(a,B)\) be the receiver's payoff given posterior belief \(\mu\) on the good state. The function \(\hat{v}:[0,1]\to\mathbb{R}\) is piecewise linear and weakly convex. It is linear on an interval \([\underline{\mu},\bar{\mu}]\) if and only if there exists an action \(a\) such that it is receiver optimal at every belief \(\mu\in[\underline{\mu},\bar{\mu}]\). The receiver's ex ante payoff in a perfect Bayesian equilibrium \((\sigma,\rho,\mu)\) is \(\sum_{m\in E}\bar{\sigma}(m)\hat{v}(\mu(m))\), where \(\bar{\sigma}(m)=\sum_{e\in E}\sigma(m|e)[F_G(e)\pi_0+F_B(e)(1-\pi_0)]\) is the probability that \(m\) is disclosed. Notice that \(\sum_{m\in E}\bar{\sigma}(m)\mu(m)=\pi_0\). By convexity, the receiver's ex ante payoff is at least \(\hat{v}(\pi_0)\).

	Let us start with a simple case. Suppose that there exists a receiver optimal perfect Bayesian equilibrium where all sender types receive the same payoff. Without loss of generality, assume that the receiver chooses the same action \(\bar{a}\) after seeing any on-path disclosure. The receiver's ex ante payoff in \(\mathcal{E}\) is therefore \(\hat{v}(\pi_0)\), and \(\bar{a}\in\phi(\pi_0)\). Hence, in any purifiable truthful equilibrium, the receiver must also get \(\hat{v}(\pi_0)\). Moreover, letting \(\mu^\star\) be the receiver's system of beliefs in purifiable truthful equilibria, \(\bar{a}\in\phi(\mu^\star(m))\) for all on-path disclosures \(m\). Hence, there exists a purifiable truthful equilibrium where the receiver always takes action \(\bar{a}\) on the equilibrium path, and it is payoff equivalent to the receiver optimal perfect Bayesian equilibrium.

	For the more general case, let \(\mathcal{E}\) be a receiver optimal perfect Bayesian equilibrium. It uniquely defines a partition \(\{E_1,E_2,\dots,E_N\}\) of the evidence space \(E\), such that the sender receives distinct equilibrium payoffs given evidence in each \(E_i\). Notice that in the equilibrium \(\mathcal{E}\), the sender's disclosure is in the same element of partition \(E_i\) as her evidence. Therefore, restricted to each \(E_i\), \(\mathcal{E}\) is a well-defined perfect Bayesian equilibrium \(\mathcal{E}_i\) of the evidence game \(\mathcal{G}_i=\langle\pi_0,(E_i,\precsim),F^i_G,F^i_B,A,u_R\rangle\), where \(F^i_\omega(\cdot)=F_\omega(\cdot)/F_\omega(E_i)\) is the distribution of evidence in state \(\omega\) conditional on \(E_i\). Moreover, \(\mathcal{E}_i\) is receiver optimal, and the receiver's ex ante payoff in \(\mathcal{E}\) is the weighted average of his ex ante payoff in each \(\mathcal{E}_i\).

	We have shown that there exists a purifiable truthful equilibrium of \(\mathcal{G}_i\) that is payoff equivalent to each \(\mathcal{E}_i\). This allows us to define a tuple \((\sigma,\rho,\mu)\) such that, restricted to each \(E_i\), it is the purifiable truthful equilibrium of \(\mathcal{G}_i\) that is payoff equivalent to \(\mathcal{E}_i\). It is easy to verify using Proposition 1 in \citet{jiang_2019}, which characterizes the unique receiver's system of beliefs in purifiable truthful equilibria, and Theorem \ref{pro2} above that \((\sigma,\rho,\mu)\) is a purifiable truthful equilibrium of the evidence game \(\mathcal{G}\). Moreover, it is payoff equivalent to \(\mathcal{E}\) by construction.

	We now show that the receiver's ex ante payoff is the same in all purifiable truthful equilibria. Hence, all purifiable truthful equilibria are receiver optimal. Let \((\sigma,\rho,\mu)\) be a purifiable truthful equilibrium. Since the sender has a strict incentive to persuade the receiver, \(\sigma(m|e)>0\) only if \(m\in\argmax_{m'\precsim e}\mu(m')\). Let \(\mu_1<\mu_2<\dots<\mu_N\) be elements of \(\mu(E)\), and define \(E_i=\{e:\max_{m\precsim e}\mu(m)=\mu_i\}\) for each \(\mu_i\). The receiver's ex ante payoff in the purifiable truthful equilibrium is then \(\sum_{i=1}^N[F_G(E_i)\pi_0+F_B(E_i)(1-\pi_0)]\hat{v}(\mu_i)\). By Theorem \ref{pro2}, \(\mu=\mu^\star\) in all purifiable truthful equilibria. Therefore, the receiver's ex ante payoff does not depend on the choice of purifiable truthful equilibrium.

	We now turn to the problem where the receiver commits to a deterministic mechanism. We make two observations. First, a revelation principle applies, and it is without loss to focus on truthful mechanisms. A mechanism \(\gamma\) is \emph{truthful} if \(\gamma:(E,\precsim)\to A\) is weakly increasing. Second, it is without loss to assume that no receiver action is dominated. The assumption of increasing differences ensures that if an action \(a_i\) is dominated, then given any posterior belief, the receiver's expected payoff from taking an adjacent action (i.e., \(a_{i-1}\) or \(a_{i+1}\)) is at least the same as taking action \(a_i\). Moreover, replacing action \(a_i\) with an adjacent action in a mechanism does not change the sender's incentive constraints. Therefore, there exists an optimal deterministic mechanism that does not involve dominated receiver actions.

	Let \(\mu_1<\mu_2<\dots<\mu_K\) be such that at each belief \(\mu_i\), \(\phi(\mu_i)=\{a_i\}\), i.e., \(a_i\) is the unique optimal action for the receiver. A deterministic mechanism \(\gamma:E\to A\) can be identified with \(\tilde{\gamma}:E\to\{\mu_i\}_{i=1}^K\) such that \(\tilde{\gamma}(m)=\mu_i\) if and only if \(\gamma(m)=a_i\). Let \(\eta^n\xrightarrow{w} 0\) be a sequence of disturbances. Recall that \(\tau(\mu,\cdot)\) is a.e. a singleton for all \(\mu\in[0,1]\). By slight abuse of notation, we shall define
	\begin{equation*}
		\varphi^n(\mu) = \int\tau(\mu,\zeta)\eta^n(d\zeta).
	\end{equation*}
	Since \(\varphi^n\) is strictly increasing, \(\varphi^n\circ\tilde{\gamma}: E\to\mathbb{R}\) is weakly increasing given any truthful mechanism \(\gamma\). Hence, in the auxiliary game \(\mathcal{G}(\varphi^n)\), it is incentive compatible for all sender types to disclose truthfully if the receiver commits to the deterministic action plan \(\varphi^n\circ\tilde{\gamma}\), and the receiver's expected payoff by making this commitment is
	\begin{equation} \label{eq.a3}
		\mathbb{E}\left[\int v_R(\tau(\tilde{\gamma}(e),\zeta),\omega|\zeta)\eta^n(d\zeta)\right],
	\end{equation}
	where the expectation is taken over the state of the world \(\omega\) and the sender's type \(e\). As \(n\to\infty\), (\ref{eq.a3}) converges to
	\begin{equation*}
		\mathbb{E}\left[u_R(\gamma(e),\omega)\right],
	\end{equation*}
	which is the receiver's expected payoff by committing to the mechanism \(\gamma\) in the original evidence game \(\mathcal{G}\). In each auxiliary game \(\mathcal{G}(\varphi^n)\), HKP shows that there is no value of committing to a deterministic mechanism, and (\ref{eq.a3}) is bounded by the receiver's truth-leaning equilibrium payoff. By the duality between truth-leaning equilibria of the auxiliary game \(\mathcal{G}(\varphi^n)\) and truth-leaning equilibria of the disturbed game \(\mathcal{G}_R(\eta^n)\), (\ref{eq.a3}) is also bounded by the receiver's truth-leaning equilibrium payoff in the disturbed game \(\mathcal{G}_R(\eta^n)\). As \(n\to\infty\), the receiver's expected payoff from committing to the mechanism \(\gamma\) is therefore bounded by his purifiable truthful equilibrium payoff in the original evidence game \(\mathcal{G}\). The receiver can achieve any equilibrium payoff using a deterministic mechanism, since every perfect Bayesian equilibrium coexists with a perfect Bayesian equilibrium that gives the receiver the same payoff and in which the receiver uses a pure strategy. Therefore, the receiver's payoff in the optimal deterministic mechanism must equal his purifiable truthful equilibrium payoff.
\end{proof}

\bibliography{Purification}

\end{document}